\documentclass[iop]{emulateapj}

\usepackage{graphicx,amsmath,natbib}
\usepackage{here}
\usepackage{epsf}
\usepackage{epstopdf}
\usepackage{xspace}

\usepackage[normalem]{ulem} % for regular text; use \sout{}
\usepackage{cancel} % for equations; use \cancel{}

\usepackage{color}

\newcommand{\um}{\ensuremath{\mu\rm{m}}\xspace}
\newcommand{\kms}{\ensuremath{\rm{km\,s}^{-1}}\xspace}
\newcommand{\alphaco}{\ensuremath{\alpha_{\rm{CO}}}\xspace}
\newcommand{\gdr}{\ensuremath{\delta_{\rm{GDR}}}\xspace}
\newcommand{\dustkappa}{\ensuremath{\kappa_{\rm{870\,\um}}}\xspace}
\newcommand{\sigsfr}{\ensuremath{\Sigma_{\rm{SFR}}}\xspace}
\newcommand{\siggas}{\ensuremath{\Sigma_{\rm{gas}}}\xspace}
\newcommand{\uJy}{\ensuremath{\mu\rm{Jy}}\xspace}

\newcommand{\lir}{\ensuremath{L_{\rm{IR}}}\xspace}
\newcommand{\msol}{\ensuremath{\rm{M}_\odot}\xspace}
\newcommand{\lsol}{\ensuremath{\rm{L}_\odot}\xspace}
\newcommand{\lprime}{\ensuremath{\rm{L}_{\rm{CO}}'}\xspace}

\newcommand{\percc}{\ensuremath{\rm{cm}^{-3}}\xspace}

\newcommand{\nht}{\ensuremath{n_\mathrm{H_2}}\xspace}
\newcommand{\tkin}{\ensuremath{T_\mathrm{kin}}\xspace}
\newcommand{\etal}{et~al.\xspace}

\def\Arizona{1}
\def\Diego{2}
\def\ESOGarching{3}
\def\Cambridge{4}
\def\KICPChicago{5}
\def\EFIChicago{6}
\def\PhysicsUChicago{7}
\def\AAUChicago{8}
\def\Dal{9}
\def\UWS{10}
\def\CSIRO{11}
\def\Davis{12}
\def\UFlorida{13}
\def\PUC{14}
\def\Stanford{15}
\def\UCLA{16}
\def\MPIfR{17}
\def\Illinois{18}

\begin{document}

\title{Sub-kiloparsec Imaging of Cool Molecular Gas in\\
Two Strongly Lensed Dusty, Star-Forming Galaxies}

%\author{SPT SMG Collaboration\altaffilmark{\SPT}}
%\altaffiltext{\SPT}{South Pole Telescope, Antarctica}

\shortauthors{J. S. Spilker, et al.}

\author{
J.~S.~Spilker$^{\Arizona}$,
M.~Aravena$^{\Diego}$,
D.~P.~Marrone$^{\Arizona}$,   
%%%%%%%%%%%%%%%
%%% EVERYONE ELSE
%%%%%%%%%%%%%%%
%J.~E.~Aguirre$^{\UPenn}$,
%M.~L.~N.~Ashby$^{\CfA}$,
%B.~A.~Benson$^{\KICPChicago,\EFIChicago}$,
M.~B\'ethermin$^{\ESOGarching}$,
%C.~M.~Bradford$^{\JPL}$,
M.~S.~Bothwell$^{\Cambridge}$,
%M.~Brodwin$^{\Miss}$,
J.~E.~Carlstrom$^{\KICPChicago,\PhysicsUChicago,\EFIChicago,\AAUChicago}$, 
S.~C.~Chapman$^{\Dal}$,
%T.~M.~Crawford$^{\KICPChicago,\AAUChicago}$, 
J.~D.~Collier$^{\UWS,\CSIRO}$,
C.~de~Breuck$^{\ESOGarching}$,
C.~D.~Fassnacht$^{\Davis}$,
T.~Galvin$^{\UWS,\CSIRO}$,
A.~H.~Gonzalez$^{\UFlorida}$, 
J.~Gonz\'{a}lez-L\'{o}pez$^{\PUC}$,
K.~Grieve$^{\UWS}$,
%T.~R.~Greve$^{\UCL}$,	
%B.~Gullberg$^{\ESOGarching}$, 
Y.~Hezaveh$^{\Stanford}$,
%G.~P.~Holder$^{\McGill}$, 
%W.~L.~Holzapfel$^{\Berkeley}$, 
%K.~Husband$^{\Bristol}$,
%R.~Keisler$^{\KICPChicago,\PhysicsUChicago}$, 
J.~Ma$^{\UFlorida}$, 
M.~Malkan$^{\UCLA}$,
%E.~J.~Murphy$^{\IPAC}$,
A.~O'Brien$^{\UWS,\CSIRO}$,
%C.~L.~Reichardt$^{\Berkeley}$, 
K.~M.~Rotermund$^{\Dal}$,
%B.~Stalder$^{\CfA}$, 
%A.~A.~Stark$^{\CfA}$, 
M.~Strandet$^{\MPIfR}$, 
J.~D.~Vieira$^{\Illinois}$,
A.~Wei\ss$^{\MPIfR}$,
%N.~Welikala$^{\Oxford}$,
G.~F.~Wong$^{\UWS,\CSIRO}$
}

\altaffiltext{\Arizona}{Steward Observatory, University of Arizona, 933 North Cherry Avenue, Tucson, AZ 85721, USA; \href{mailto:jspilker@as.arizona.edu}{jspilker@as.arizona.edu}}
%\altaffiltext{\ESOChile}{European Southern Observatory, , Alonso de Cordova 3107, Casilla 19001 Vitacura Santiago, Chile.}
\altaffiltext{\Diego}{N\'ucleo de Astronom\'{\i}a, Facultad de Ingenier\'{\i}a, Universidad Diego Portales, Av. Ej\'ercito 441, Santiago, Chile}
\altaffiltext{\ESOGarching}{European Southern Observatory, Karl Schwarzschild Stra\ss e 2, 85748 Garching, Germany}
%\altaffiltext{\UPenn}{University of Pennsylvania, 209 South 33rd Street, Philadelphia, PA 19104, USA}
%\altaffiltext{\CfA}{Harvard-Smithsonian Center for Astrophysics, 60 Garden Street, Cambridge, MA 02138, USA}
%\altaffiltext{\JPL}{Jet Propulsion Laboratory, 4800 Oak Grove Drive, Pasadena, CA 91109, USA}
\altaffiltext{\Cambridge}{Cavendish Laboratory, University of Cambridge, JJ Thompson Ave, Cambridge CB3 0HA, UK}
%\altaffiltext{\Miss}{Department of Physics and Astronomy, University of Missouri, 5110 Rockhill Road, Kansas City, MO 64110, USA}
\altaffiltext{\KICPChicago}{Kavli Institute for Cosmological Physics, University of Chicago, 5640 South Ellis Avenue, Chicago, IL 60637, USA}
\altaffiltext{\EFIChicago}{Enrico Fermi Institute, University of Chicago, 5640 South Ellis Avenue, Chicago, IL 60637, USA}
\altaffiltext{\PhysicsUChicago}{Department of Physics, University of Chicago, 5640 South Ellis Avenue, Chicago, IL 60637, USA}
\altaffiltext{\AAUChicago}{Department of Astronomy and Astrophysics, University of Chicago, 5640 South Ellis Avenue, Chicago, IL 60637, USA}
%\altaffiltext{\Argonne}{Argonne National Laboratory, Argonne, IL 60439, USA}
\altaffiltext{\Dal}{Dalhousie University, Halifax, Nova Scotia, Canada}
\altaffiltext{\UWS}{University of Western Sydney, Locked Bag 1797, Penrith, NSW 2751, Australia}
\altaffiltext{\CSIRO}{CSIRO Astronomy \& Space Science, Australia Telescope National Facility, PO Box 76, Epping, NSW 2121, Australia}
\altaffiltext{\Davis}{Department of Physics,  University of California, One Shields Avenue, Davis, CA 95616, USA}
\altaffiltext{\UFlorida}{Department of Astronomy, University of Florida, Gainesville, FL 32611, USA}
\altaffiltext{\PUC}{Instituto de Astrof\'{i}sica, Facultad de F\'{i}sica, Pontificia Universidad Cat\'{o}lica de Chile, Av. Vicu\~{n}a Mackenna 4860, 782-0436 Macul, Santiago, Chile}
%\altaffiltext{\UCL}{Department of Physics and Astronomy, University College London, Gower Street, London WC1E 6BT, UK}
\altaffiltext{\Stanford}{Kavli Institute for Particle Astrophysics and Cosmology, Stanford University, Stanford, CA 94305, USA}
%\altaffiltext{\Berkeley}{Department of Physics, University of California, Berkeley, CA 94720, USA}
%\altaffiltext{\Bristol}{H.H. Wills Physics Laboratory, University of Bristol, Tyndall Avenue, Bristol BS8 1TL, UK}
\altaffiltext{\UCLA}{Department of Physics and Astronomy, University of California, Los Angeles, CA 90095-1547, USA}
%\altaffiltext{\IPAC}{Infrared Processing and Analysis Center, California Institute of Technology, MC 220-6, Pasadena, CA 91125, USA}
\altaffiltext{\MPIfR}{Max-Planck-Institut f\"{u}r Radioastronomie, Auf dem H\"{u}gel 69 D-53121 Bonn, Germany}
\altaffiltext{\Illinois}{Department of Astronomy and Department of Physics, University of Illinois, 1002 West Green St., Urbana, IL 61801}
%\altaffiltext{\Oxford}{Department of Physics, Oxford University, Denis Wilkinson Building, Keble Road, Oxford, OX1 3RH, UK}

%%%%%%%%%%%%%%%%%%%%%%%%%%%%%%%%%%%%%%%%%%%%%%%%%%%%%%%%%%%%%%%%%%%%%%%%%%%%%%%%%%%%%
%%%%%%%%%%%%%%%%%%%%%%%%%%%%%%%%%%%% ABSTRACT %%%%%%%%%%%%%%%%%%%%%%%%%%%%%%%%%%%%%%%
%%%%%%%%%%%%%%%%%%%%%%%%%%%%%%%%%%%%%%%%%%%%%%%%%%%%%%%%%%%%%%%%%%%%%%%%%%%%%%%%%%%%%
\begin{abstract}

We present spatially-resolved imaging obtained with the Australia Telescope
Compact Array (ATCA) of three CO lines in two high-redshift gravitationally 
lensed dusty star-forming galaxies, discovered by the South Pole Telescope. 
Strong lensing allows us to probe the structure and dynamics
of the molecular gas in these two objects, at $z=2.78$ and $z=5.66$, with effective 
source-plane resolution of less than 1\,kpc.  We model the lensed emission from multiple CO
transitions and the dust continuum in a consistent manner, finding that the cold molecular
gas as traced
by low-$J$ CO always has a larger half-light radius than the 870\,\um
dust continuum emission. This size difference leads to up to 50\%
differences in the magnification factor for the cold gas compared to dust.  In 
the z=2.78 galaxy, 
these CO observations confirm that the background
source is undergoing a major merger, while the velocity field
of the other source is more complex.  We use the ATCA CO observations 
and comparable resolution Atacama Large Millimeter/submillimeter Array 
dust continuum imaging of the
same objects to constrain the CO-H$_2$ conversion factor with three different 
procedures, finding good agreement between the methods and values consistent with those
found for rapidly star-forming systems.  We discuss these galaxies in the 
context of the star formation -- gas mass surface density relation, noting that the change in emitting
area with observed CO transition must be accounted for when comparing high-redshift 
galaxies to their lower redshift counterparts.

\end{abstract}


%%%%%%%%%%%%%%%%%%%%%%%%%%%%%%%%%%%%%%%%%%%%%%%%%%%%%%%%%%%%%%%%%%%%%%%%%%%%%%%%%%%%%
%%%%%%%%%%%%%%%%%%%%%%%%%%%%%%%%%% Introduction %%%%%%%%%%%%%%%%%%%%%%%%%%%%%%%%%%%%%
%%%%%%%%%%%%%%%%%%%%%%%%%%%%%%%%%%%%%%%%%%%%%%%%%%%%%%%%%%%%%%%%%%%%%%%%%%%%%%%%%%%%%
\section{Introduction} \label{intro}

Carbon monoxide ($^{12}\rm{C}^{16}\rm{O}$; hereafter CO) has long been known as a
tracer of molecular hydrogen gas in galaxies.  Molecular gas is the fuel for new
generations of stars (for recent reviews, see
\citealt{bolatto13} and \citealt{carilli13}), so accurately diagnosing its
abundance, kinematics, and morphology can shed light on the astrophysics of star
formation.
The most intense bouts of star formation in the universe appear to occur in 
dusty, star-forming galaxies (DSFGs) at high redshift \citep[e.g.,][]{casey14}.  
These galaxies are heavily enshrouded in dust, which absorbs the ultraviolet radiation
from massive young stars and reradiates at far-IR and submillimeter wavelengths.
These galaxies lie in contrast to the bulk of the high-redshift galaxy
population, which form stars more slowly in less dusty, generally isolated 
systems \citep[e.g.,][]{forsterschreiber09,daddi10,tacconi13}. Such ``normal'' 
galaxies are selected by their stellar, rather than dust, emission, which generally 
excludes the highly obscured DSFG population.
Together with
rapid star formation, DSFGs contain comparably large reservoirs of molecular
gas \citep[$>10^{10}$\,\msol; e.g.,][]{greve05,ivison11,bothwell13} that make up a
significant fraction ($\sim$20--80\%)
of the total baryonic mass \citep[e.g.,][]{carilli10,ivison11,carilli13}.

The most extreme DSFGs are likely to be galaxies undergoing major mergers
(e.g., \citealt{narayanan10,hayward12,fu13,ivison13}, though see 
\citealt{carilli10,hodge12,hodge15} for a notable counterexample),
with star formation rates (SFRs) enhanced
by gas being funneled to the center of the system after being disrupted
during the collision.  The merger kinematically manifests as a disordered
velocity field or multiple components closely separated in position and/or velocity
\citep[e.g.,][]{engel10,fu13}.  Such an extreme level of star formation can
likely only be sustained for a period of $\lesssim100$\,Myr \citep{greve05}, and
thus the brightest DSFGs are also relatively rare.

Extensive effort has gone towards studying gas and dust both in the local
universe and at high redshift in order to understand the physics and history of star 
formation.  One of the most studied correlations is the power-law relationship
between the gas surface density, \siggas, and the SFR surface density, \sigsfr
(\citealt{schmidt59,kennicutt98b}; see \citealt{kennicutt12} for a recent review).
On scales of a few hundred parsecs, the two quantities appear linearly related
\citep[e.g.,][]{schruba11,leroy13}, though power-law exponents ranging from
sub-linear to quadratic have also been theoretically predicted and observationally
confirmed depending on methodology \citep[e.g.,][]{krumholz09,liu11,
fauchergiguere13,shetty13}.  This star formation (SF) relation
(or Schmidt-Kennicutt relation)
is one ingredient in many theoretical prescriptions for star formation, so
understanding its mathematical form and range of applicability is important
for understanding the buildup of stellar mass.

The steps to derive a molecular gas mass from the luminosity of a low-$J$ CO transition
are not straightforward, and a variety of techniques have been presented
in the literature \citep[e.g.,][]{bolatto13}.  The
conversion factor, \alphaco, varies with the kinematic state of the gas (through the
escape fraction of CO photons) and the gas metallicity (through CO formation
and destruction processes).  A variety of observations
suggest that a value of $\alphaco \sim 3.6-4.5$\,\msol\,pc$^{-2}$\,(K\,\kms)$^{-1}$ 
(including a 36\% mass contribution from the cosmological abundance of helium; hereafter we suppress the units of \alphaco) is
applicable to the Milky Way and nearby quiescently star-forming galaxies with
approximately solar metallicity \citep[e.g.,][]{solomon87,abdo10,sandstrom13}.  In regions
of vigorous star formation, however, \alphaco decreases by a factor of several
\citep[e.g.,][]{downes98,tacconi08}.

As a further complication, the high-\sigsfr galaxies that are much more 
common at high redshift are rare in the local universe, which makes their 
exploration more difficult. Due to the faintness of the 
lowest transitions of CO, most high-redshift studies
of molecular gas have used either spatially unresolved observations, or brighter, 
higher-$J$ transitions with higher excitation conditions than the ground
state, or in some cases both.  An additional conversion from the observed CO transition
to CO(1--0) is required, which depends on the temperature, density, and structure of the 
interstellar medium (ISM).
Resolved observations of the lowest CO transitions are needed to test the SF
relation on sub-galactic scales.

Such high-resolution studies are aided by the use of gravitational lensing, in which
a background object is magnified by a foreground structure, usually a massive
elliptical galaxy or galaxy cluster.  For example, \citet{rawle14} use 
high-resolution maps of dust continuum emission and [CII] and CO(1-0) emission
to spatially and spectrally decompose a source at $z=5.2$ predominantly lensed
by a $z=0.63$ galaxy, finding variations in the efficiency of star formation
of a factor of $\sim$6$\times$ within a 4\,kpc region in the source plane.
\citet{thomson15} probe $\sim100$\,pc scales in the $z=2.3$ galaxy 
SMM\,J2135-0102 (``the Eyelash''; \citet{swinbank10}). These authors studied
the Schmidt-Kennicutt relation in individual star-forming clumps in this galaxy,
and found evidence that the clumps are offset towards higher star formation
efficiency compared to the galaxy as a whole. 

Bright lensed galaxies are rare, but
recent large surveys conducted by the South Pole Telsecope (SPT; 
\citealt{carlstrom11,vieira10,mocanu13}) and \textit{Herschel}
\citep{negrello10,wardlow13} have discovered large numbers of lensed DSFGs.
Subsequent spectroscopy and high-resolution imaging have confirmed that the
large majority of these objects indeed lie at high redshifts and are
lensed \citep{weiss13,vieira13,hezaveh13,harris12,bussmann13}.  In particular, 
\citet{weiss13} used the Atacama Large Millimeter/submillimeter Array (ALMA)
to conduct a redshift survey of 26 DSFGs discovered by the SPT, finding a
median redshift $\langle z \rangle = 3.5$.  Additionally, \citet{aravena13} and
Aravena et~al., \textit{in prep.} surveyed 18 of these galaxies in low-$J$ CO (either
CO(1--0) or CO(2--1)).

In this paper, we present high-resolution observations of low-$J$ CO emission
in two DSFGs from the SPT sample performed with the Australia Telescope
Compact Array (ATCA).  Both objects have been observed at comparable ($\sim$0.5'')
resolution by ALMA at 870\,\um, with lens models determined from these
data \citep{hezaveh13}.  SPT-S J053816-5030.8 (SPT0538-50), at $z=2.78$, 
is representative of the typical DSFG population in redshift, 870\,\um flux
density, and dust temperature. This object was studied in detail 
by \citet{bothwell13b}, who
showed evidence for two velocity components in CO(7--6) separated by 
$\Delta v \sim 350$\,\kms.
Intriguingly, the lens model of this source \citep{hezaveh13} also 
required two dust components to reproduce the ALMA data,
suggesting a possible physical connection between the velocity structure
and the continuum structure.
SPT-S J034640-5204.9 (SPT0346-52), at $z=5.66$, is among the highest-redshift
DSFGs known. The ALMA lens model indicates that it is also the most intrinsically
luminous object in the SPT sample, and its \sigsfr approaches or surpasses the
Eddington limit for radiation pressure on dust grains \citep{thompson05}. While
not representative of the typical DSFG in the SPT sample, SPT0346-52 allows us
to study the conditions of the ISM at their most extreme.

The layout of this paper is as follows.  In Section~\ref{obs}, we describe the 
ATCA 7\,mm and 3\,mm observations of CO lines in these two objects.  In 
Section~\ref{lensmodels}, we describe our procedure for modeling the effects
of gravitational lensing in both the ATCA and ALMA data.  The morphological
and kinematic results of these
lens models are given in Section~\ref{results}.  In Section~\ref{discussion},
we use the lens modeling results to discuss the effects of preferential
source magnification, determine the \alphaco factor in each source, and 
place these sources in the context of the SF relation.  We conclude in 
Section~\ref{conclusions}.  We adopt the WMAP9 $\Lambda$CDM cosmology,
with ($\Omega_m$, $\Omega_\Lambda$, $H_0$) = (0.286, 0.713, 69.3\,\kms\,Mpc$^{-1}$)
\citep{hinshaw13}.  Throughout, we define the total infrared luminosity, \lir,
to be integrated over rest-frame 8--1000\,\um, and assume a \citet{chabrier03} initial
mass function.

%%%%%%%%%%%%%%%%%%%%%%%%%%%%%%%%%%%%%%%%%%%%%%%%%%%%%%%%%%%%%%%%%%%%%%%%%%%%%%%%%%%%%
%%%%%%%%%%%%%%%%%%%%%%%%%%%%%%%%%% Observations %%%%%%%%%%%%%%%%%%%%%%%%%%%%%%%%%%%%%
%%%%%%%%%%%%%%%%%%%%%%%%%%%%%%%%%%%%%%%%%%%%%%%%%%%%%%%%%%%%%%%%%%%%%%%%%%%%%%%%%%%%%
\section{Observations} \label{obs}

%-----------------------------------------------------------------------------------
%%%%%%%%%%%%%%%%%%%%%%%%%%% FIGURE 1: SPT0346-52 DATA %%%%%%%%%%%%%%%%%%%%%%%%%%%%%%
%-----------------------------------------------------------------------------------
\begin{figure*}[htb]%
\centering
\includegraphics[width=\textwidth]{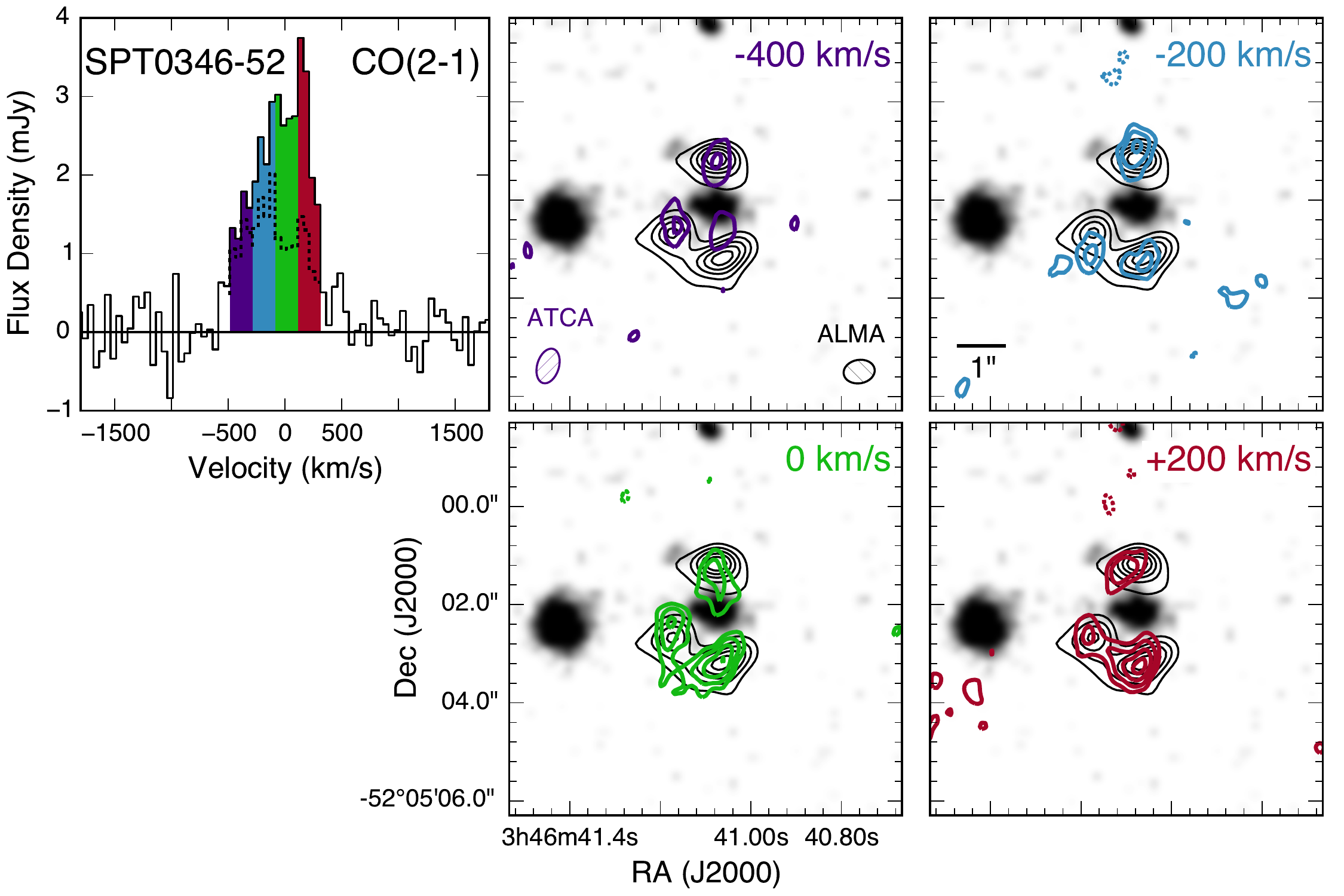}%
\caption{
Spectrum and channel maps of CO(2--1) observed in SPT0346-52.  \textit{Left:} 
Integrated spectrum derived from short, unresolved baselines, as presented
in Aravena et~al., \textit{in prep.} The four 200\,\kms channels we model are colored.
The dashed line shows the intrinsic spectrum using the magnification factors in 
Table~\ref{tab:sourceprops}, multiplied by 4$\times$ for clarity.
\textit{Right panels:} Channel maps of the high-resolution ATCA CO(2--1) observations
presented here, colored as in the left panel.  The greyscale image of
the lens galaxy is from co-added \textit{HST}/WFC3 F140W + F160W
images \citep{vieira13}. Thin black 
contours show the ALMA 870\,\um dust continuum images.  The ATCA images are shown
in steps of 2$\sigma$ starting at $\pm$3$\sigma$
(1$\sigma = 10.6$\,mJy\,\kms\,beam$^{-1}$).  The ALMA images
are shown in steps of 5$\sigma$ (1$\sigma = 0.80$\,mJy\,beam$^{-1}$). Both
datasets reach a resolution of approximately 0.45''$\times$0.65''.
}%
\label{fig:images0346}%
\end{figure*}
%-----------------------------------------------------------------------------------

%-----------------------------------------------------------------------------------
%%%%%%%%%%%%%%%%%%%%%%%%%%% FIGURE 2: SPT0538-50 DATA %%%%%%%%%%%%%%%%%%%%%%%%%%%%%%
%-----------------------------------------------------------------------------------
\begin{figure*}[htb]%
\centering
\includegraphics[width=\textwidth]{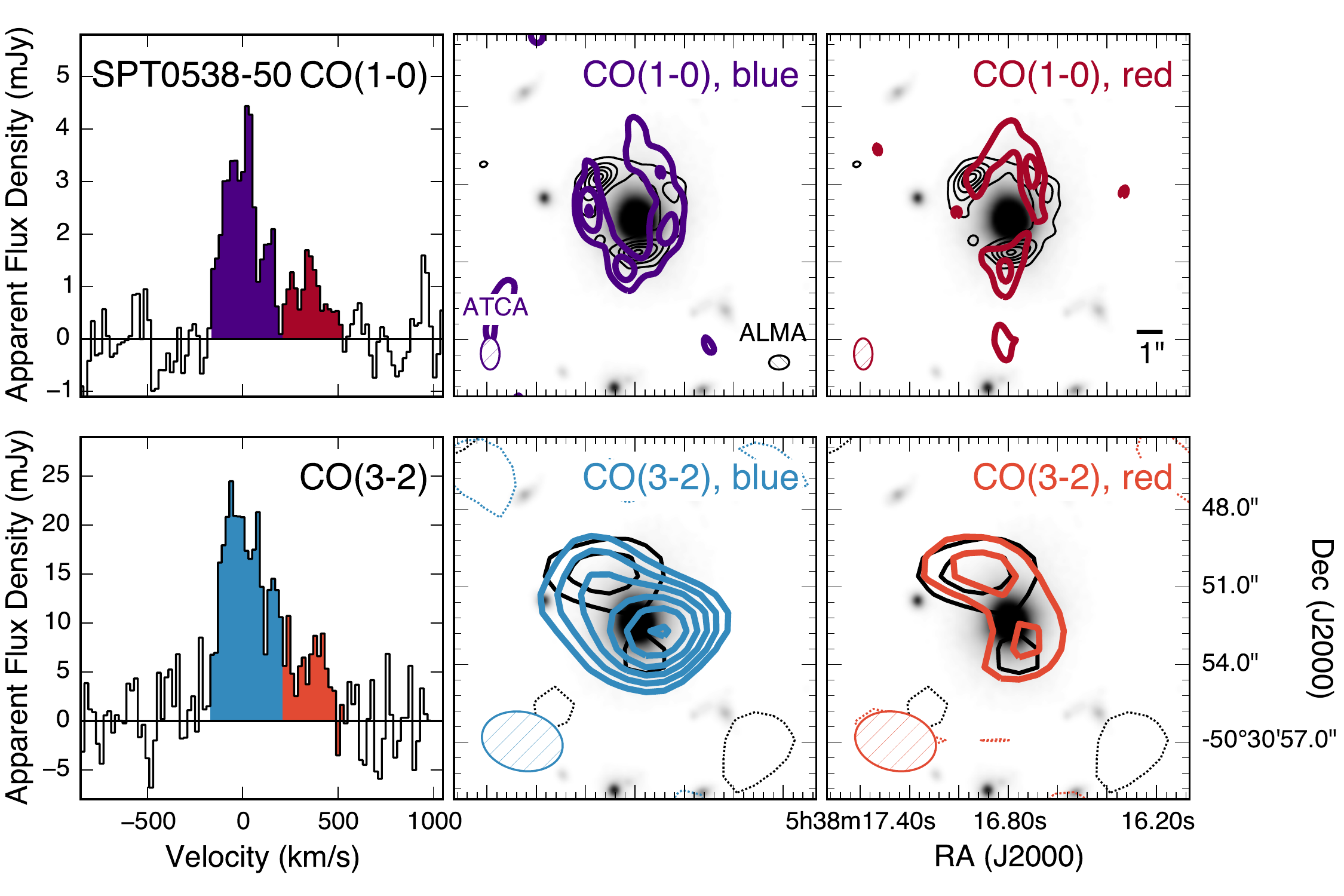}%
\caption{
Spectra and channel maps of CO(1--0) and CO(3--2) observed by ATCA in SPT0538-50.
\textit{Left panels:} Integrated spectra of each line derived from short
baselines in which the source is unresolved. We separate the two velocity components
seen in these and other spectral lines as in \citet{bothwell13b}.
\textit{Top center and right panels:} Channel maps of CO(1--0) of each of the 
colored velocity bins in the
top left panel. The greyscale image is a co-added \textit{HST}/WFC3 F140W + 
F160W image. Also shown
as thin black contours are the ALMA 870\,\um dust continuum observations.  ATCA
contours are shown in steps of 2$\sigma$, starting at $\pm$3$\sigma$
(1$\sigma = 12.7$\,mJy\,
\kms\,beam$^{-1}$), with a resolution of approximately 0.7''$\times$1.2''.  
The red velocity component
is not significantly detected at the depth and resolution of these observations. ALMA
contours are shown in steps of 10$\sigma$ starting at +5$\sigma$ (1$\sigma \sim 0.46$\,
mJy\,beam$^{-1}$).  
\textit{Bottom center and right panels:} Channel maps of CO(3--2) in each of the 
colored channels in the lower left panel. Thick black contours show the 3.3\,mm continuum
emission from the same data.  All ATCA contours in these panels are shown in 
steps of 2$\sigma$ starting at $\pm$3$\sigma$. For the CO(3--2) data, 1$\sigma = 145$
\,mJy\,\kms\,beam$^{-1}$; for the 3.3\,mm continuum data, 1$\sigma \sim 70$
\,$\mu$Jy\,beam$^{-1}$.  The ATCA data have a resolution of approximately
2.2''$\times$3.3''. It is clear from this figure that the red CO(3--2)
component is morphologically similar to the dust emission at the same wavelength,
dominated by the bright, compact dust component found in the ALMA lens models
(see text).
}%
\label{fig:images0538}%
\end{figure*}
%-----------------------------------------------------------------------------------

%%%%%%%%%%%%%%%%%%%%%%%%%%%%%%%%%%%%%%%%%%%%%%%%%%%%%%%%%%%%%%%
%%%%%%%%%%%%%%%%%% Observing Summary Table %%%%%%%%%%%%%%%%%%%%
%%%%%%%%%%%%%%%%%%%%%%%%%%%%%%%%%%%%%%%%%%%%%%%%%%%%%%%%%%%%%%%
\begin{deluxetable*}{llclcccc}
\tablecaption{Observational Summary \label{tab:obstable}}
\startdata
\tableline
\\
Source & Line & $\nu_\mathrm{obs}$ (GHz) & Dates & Array Configuration & Time On-Source (h) & Beam Size$^\mathrm{a}$ & $\sigma^\mathrm{b}$ ($\mu$Jy) \\
\tableline
\\
SPT0346-52 & CO(2--1) & 34.636 & 2012 10/04--10/09 & H\,214   & 5.5   & 6.4''$\times$4.8'' & 142 \\
           &          &        & 2014 01/17--01/18 & 1.5\,B   & 6.0   & 1.0''$\times$0.6'' & 180 \\
           &          &        & 2014 05/10--05/17 & 1.5\,D   & 7.2   & 1.6''$\times$0.8'' & 178 \\
           &          &        & 2013 10/23--11/08 & 6\,A  & 26.1  & 0.5''$\times$0.3'' & 66  \\
SPT0538-50 & CO(1--0) & 30.450 & 2012 07/28--07/29 & H\,75    & 8.5   & 19''$\times$13''   & 163 \\
           &          &        & 2014 01/18--01/20 & 1.5\,B   & 11.9  & 0.9''$\times$0.5'' & 119 \\
           &          &        & 2014 05/15--05/17 & 1.5\,D   & 11.3  & 1.2''$\times$0.6'' & 96  \\
           &          &        & 2013 10/24--11/08 & 6\,A  & 45.7  & 0.5''$\times$0.3'' & 42  \\
SPT0538-50 & CO(3--2) & 91.345 & 2013 08/20    & H\,168   & 8.7   & 3.1''$\times$2.2'' & 420 \\
\enddata
\tablenotetext{a}{Beam size for a naturally-weighted image.}
\tablenotetext{b}{rms sensitivity in 200\,\kms (SPT0346-52) or 350\,\kms (SPT0538-50) channels.}
\end{deluxetable*}

Table \ref{tab:obstable} summarizes the ATCA observations of CO lines in SPT0346-52
and SPT0538-50.  Further details of these observations are given below.  We also
make use of ALMA 870\,\um continuum observations of these objects, 
described further in \citet{hezaveh13}.

\subsection{ATCA Observations: 7\,mm Band}
SPT0538-50 and SPT0346-52 were observed with the ATCA 7\,mm receivers in CO(1--0) 
and CO(2--1) at observed frequencies of 30.45 and 34.64\,GHz, respectively, as
part of project IDs C2892 and C2983.  The 
sources were observed using a 6-km extended array configuration over the course of
12 nights in 2013 October-November and a compact 1.5-km array configuration over
6 nights in 2014 January and May.  The two CO lines are redshifted to similar
frequencies and can be observed without retuning the two 1\,GHz-wide basebands
available using the Compact Array Broadband Backend (CABB), allowing bandpass and
absolute flux calibration to be shared between the two sources observed in a single
track.  The bright quasars PKS1921-293 and PKS0537-441 were observed for bandpass calibration,
while the quasars PKS0322-403 and PKS0537-441 served as complex gain calibrators for SPT0346-52
and SPT0538-50, respectively.  For most tracks, the quasar PKS1934-638 was observed for
flux calibration; when this source was not available, the flux level of the bandpass
and gain calibrators from adjoining observing dates was used to set the amplitude scale.
Repeated observations of amplitude calibration sources indicate that the absolute 
flux scale is accurate to within 10\%.

The two SPT DSFGs were also observed by \citet{aravena13} and Aravena \etal, 
\textit{in prep.}
using the compact ATCA configurations H75 (SPT0538-50, project C2655) and H214
(SPT0346-52, project C2744) in 2012 July and October, respectively. 
For our present purposes,
the baselines provided by these compact array configurations provide sensitivity to
extended emission and an estimate of the total flux.  The integrated spectrum
of each source derived from these data are shown in Figs.~\ref{fig:images0346}
and \ref{fig:images0538}.  The spectrum
of SPT0538-50 has two peaks, as noted by \citet{aravena13}, which we 
discuss further in Section \ref{lensmodels}.

The three array configurations provide good $uv$ coverage on baselines from 
$\sim$100\,m -- 6\,km, and were combined 
and inverted using natural
weighting.  We ensure proper normalization of the noise levels of each dataset
by differencing successive pairs of visibilities on the same baseline and polarization.
Naturally-weighted channel maps of each galaxy are shown in  Fig.~\ref{fig:images0346}
and the upper panels of Fig.~\ref{fig:images0538}.
We show
channel maps of CO(2--1) in SPT0346-52 in 200\,\kms-wide channels, 
and separate the CO(1--0) line
of SPT0538-50 into the red and blue velocity components seen in the integrated spectrum 
(two channels, approximately 350\,\kms wide).  Note that the red velocity component of 
SPT0538-50 is not significantly detected, as the weak line flux is spread over several
synthesized beams.  The 1$\sigma$ sensitivities of these maps are 
54\,$\mu$Jy beam$^{-1}$ (170\,mK) per
200\,\kms channel in SPT0346-52 and 36\,$\mu$Jy beam$^{-1}$ (53\,mK) 
per 350\,\kms channel in SPT0538-50.

\subsection{ATCA Observations: 3\,mm Band}
For SPT0538-50, we also observed the CO(3--2) line, redshifted to 91.35\,GHz, using the
hybrid H168 array configuration on 2013 August 20 in project C2816.
The other 1\,GHz-wide baseband was
tuned to 94\,GHz.  The quasar PKS0537-441 was again used
for bandpass and complex gain calibration, while Uranus was observed for flux calibration.
The absolute flux scale at 3\,mm is expected to be accurate to within $\sim$15\%, again inferred from 
repeated observations of amplitude calibration sources.  

The data were continuum-subtracted and imaged using natural weighting
to maximize sensitivity to weak emission, giving a 
synthesized beam of $\sim$3.1$\times$
2.2''.  This resolution is sufficient to marginally resolve the source, as seen in the
bottom panels of Fig.~\ref{fig:images0538}, where we have again imaged the line in each
of the two CO velocity peaks separately.  These maps reach a sensitivity of 420\,$\mu$Jy
beam$^{-1}$ (8.6\,mK) in each 350\,\kms channel.
Due to the larger synthesized beam size
and higher line flux of the CO(3--2) line
compared to the CO(1--0) line, we are also able to detect the weak, red velocity component
in these data, at $\sim 6\sigma$ significance.

We also significantly detect the dust continuum emission at 3.3\,mm using the 
line-free channels of both
basebands, reaching a sensitivity of 90\,$\mu$Jy beam$^{-1}$.  The dust continuum emission
closely resembles the emission from the red portion of the CO(3--2) line, which we
discuss further in Section \ref{res0538}.

%%%%%%%%%%%%%%%%%%%%%%%%%%%%%%%%%%%%%%%%%%%%%%%%%%%%%%%%%%%%%%%%%%%%%%%%%%%%%%%%%%%%%
%%%%%%%%%%%%%%%%%%%%%%%%%%%%%%%%%% Lens Models %%%%%%%%%%%%%%%%%%%%%%%%%%%%%%%%%%%%%%
%%%%%%%%%%%%%%%%%%%%%%%%%%%%%%%%%%%%%%%%%%%%%%%%%%%%%%%%%%%%%%%%%%%%%%%%%%%%%%%%%%%%%
\section{Lens Modeling} \label{lensmodels}

To derive the intrinsic gas and dust properties of the two DSFGs presented
here, we must quantify the effects of gravitational lensing.  Our lens modeling 
procedure follows that described by \citet{hezaveh13}. Briefly, the lens mass profile
is represented by a Singular Isothermal Ellipsoid (SIE). For SPT0346-52,
the model also strongly favors the existence of an external shear component whose
axis is aligned with another galaxy $\sim$3'' east of the primary lens. For both sources,
the ALMA 870\,\um data are of much higher significance than the ATCA observations
presented here, so we use the best-fit lens properties derived from the ALMA
continuum data to model the source-plane in the ATCA CO data.  

The lensed CO source is represented by a parameterized model consisting of one (SPT0346-52)
or two (SPT0538-50; see below) symmetric Gaussian light profiles in each modeled velocity
channel.  Each profile has
up to four free parameters, namely, the two-dimensional centroid of the source and its
intrinsic flux and size.  While this source-plane model is undoubtedly overly
simplistic, it allows the derived properties of each velocity bin to be compared in a 
straightforward manner.  Using a parametric model additionally avoids over-fitting
the data using a large number of free parameters, as in most pixel-based
reconstruction techniques, which are more appropriate for very high resolution, very high
S/N observations.

ATCA and ALMA both
measure the Fourier components (visibilities) of the sky 
at the two-dimensional spatial frequencies
defined by pairs of antennas.  Rather than comparing to reconstructed images, where
there are strong correlations between pixels, we fit lensing models directly to
the measured visibilities.
As in \citet{hezaveh13}, we use a Bayesian Markov Chain Monte Carlo (MCMC) fitting 
procedure.  At each MCMC step, we generate a model lensed image from a given set of 
source parameters.  We then invert this image to the Fourier domain and interpolate
the model visibilities to the measured $uv$ coordinates of the ATCA data, using the
$\chi^2$ metric to determine the quality of the fit.

Lens modeling of the 870\,\um dust continuum emission of both sources was previously
presented in \citet{hezaveh13} using ALMA data with approximately 1.5''
resolution.  In the present work, we additionally make use of higher resolution
data taken as part of the same ALMA program, but which were not yet available at the
time of publication of Hezaveh \etal.  These new data were taken in an extended
array configuration available in ALMA Cycle 0, and reach $\sim$0.5'' resolution.  
Further observational details
of this ALMA program are given in \citet{hezaveh13}, while the updated 870\,\um lens
models will be presented in full in Spilker \etal, \textit{in prep}.  
For both sources, the updated
lens models using the higher-resolution data are qualitatively and quantitatively
similar to those derived by Hezaveh \etal  Source properties relevant to this work
are summarized in Table~\ref{tab:sourceprops}. 

In SPT0346-52, the background source
is well-fit by a single elliptical S\'{e}rsic light profile with a
circularized half-light radius of $610\pm30$\,pc.  The source's 
intrinsic 870\,\um flux density of $19.6\pm0.5$\,mJy is
magnified by a factor of $\mu = 5.5\pm0.1$.  

In SPT0538-50, the updated lens model requires
two source-plane components to fit the data, as in Hezaveh \etal  The source consists of
a faint, diffuse dust component of intrinsic flux density
$1.5\pm0.2$\,mJy, half-light radius $1.25\pm0.14$\,kpc
magnified by $\mu = 23.1\pm2.2$, and a brighter, more compact component 
of flux density $3.7\pm0.4$\,mJy and 
half-light radius $470\pm50$\,pc magnified by $\mu = 18.9\pm2.2$.

%%%%%%%%%%%%%%%%%%%%%%%%%%%%%%%%%%%%%%%%%%%%%%%%%%%%%%%%%%%%%%%%%%%%%%%%%%%%%%%%%%%%%
%%%%%%%%%%%%%%%%%%%%%%%%%%%%%%%%%%%% Results %%%%%%%%%%%%%%%%%%%%%%%%%%%%%%%%%%%%%%%%
%%%%%%%%%%%%%%%%%%%%%%%%%%%%%%%%%%%%%%%%%%%%%%%%%%%%%%%%%%%%%%%%%%%%%%%%%%%%%%%%%%%%%
\section{Results} \label{results}

%%%%%%%%%%%%%%%%%%%%%%%%%%%%%%%%%%%%%%%%%%%%%%%%%%%%%%%%%%%%%%%
%%%%%%%%%%%%%%% Table of Lens Modeling Results %%%%%%%%%%%%%%%%
%%%%%%%%%%%%%%%%%%%%%%%%%%%%%%%%%%%%%%%%%%%%%%%%%%%%%%%%%%%%%%%
\begin{deluxetable*}{llccccc}
\tablecaption{CO and Dust Lens Modeling Results \label{tab:sourceprops}}
\startdata
\tableline
\\
\multicolumn{7}{c}{CO Lens Model Properties}\\
Source & Line & Component & $\mu_{\mathrm{CO}}$ & \lprime & r$_\mathrm{eff, CO}$ & $\log(\siggas)$$^{\mathrm{ab}}$ \\
 & & & & (10$^{10}$\,K\,\kms\,pc$^2$) & (kpc) & (\msol\,pc$^{-2}$) \\
\tableline
\\
SPT0346-52    & CO(2-1) & -400\,\kms & 5.0 $\pm$ 0.6  & 1.6 $\pm$ 0.2 & 0.70 $\pm$ 0.13 & 3.82 $\pm$ 0.32 \\
              &         & -200\,\kms & 5.8 $\pm$ 0.7  & 2.3 $\pm$ 0.3 & 1.86 $\pm$ 0.28 & 3.14 $\pm$ 0.27 \\
              &         & 0\,\kms    & 10.0 $\pm$ 0.1 & 1.5 $\pm$ 0.1 & 0.77 $\pm$ 0.07 & 3.72 $\pm$ 0.26 \\
              &         & +200\,\kms & 10.1 $\pm$ 0.3 & 1.4 $\pm$ 0.1 & 0.73 $\pm$ 0.09 & 3.74 $\pm$ 0.27 \\
SPT0538-50\,B & CO(1-0) & blue     & 15.7 $\pm$ 2.3 & 1.7 $\pm$ 0.3 & 2.34 $\pm$ 0.43 & 2.86 $\pm$ 0.26 \\
SPT0538-50\,A &         & red      &        --      &       --      &        --       &  --  \\
SPT0538-50\,B & CO(3-2) & blue     & 25.7 $\pm$ 1.2 & 0.94 $\pm$ 0.13 & 1.03 $\pm$ 0.14 & -- \\
SPT0538-50\,A &         & red      & 22.6 $\pm$ 1.2 & 0.51 $\pm$ 0.07 & 0.68 $\pm$ 0.12 & -- \\
\\
\tableline
\\
\multicolumn{7}{c}{Dust Lens Model Properties}\\
Source & Component & $\mu_{870\,\um}$ & S$_{870\,\um}$ & \lir & r$_\mathrm{eff, 870\,\um}$ & $\log(\sigsfr)$$^\mathrm{b}$ \\
 & & & (mJy) & 10$^{12}$\,\lsol & (kpc) & (\msol\,yr$^{-1}$\,kpc$^{-2}$) \\
\tableline
\\
SPT0346-52    &         &  5.5 $\pm$ 0.1 & 19.6 $\pm$ 0.5 & 36.3 $\pm$ 5.4 & 0.61 $\pm$ 0.03 & 3.19 $\pm$ 0.08 \\
SPT0538-50\,A &  bright & 18.9 $\pm$ 2.2 &  3.7 $\pm$ 0.4 & 3.51 $\pm$ 0.50& 0.47 $\pm$ 0.05 & 2.40 $\pm$ 0.11 \\
SPT0538-50\,B &  faint  & 23.1 $\pm$ 2.2 &  1.5 $\pm$ 0.2 & 1.48 $\pm$ 0.24& 1.25 $\pm$ 0.14 & 1.18 $\pm$ 0.12 \\
\enddata
\tablecomments{When modeling the CO emission of both objects, we fix the parameters of the lens to their best-fit values derived from the ALMA data. For SPT0538-50, we also fix the location of the A and B source components to the positions derived from the ALMA data.}
\tablenotetext{a}{Assuming the average CO-H$_2$ conversion factors determined in Section~\ref{alphaco},1.4 for SPT0346-52 and 1.5 for SPT0538-50. We propagate an uncertainty of 50\% into the determinations of \siggas.}
\tablenotetext{b}{Surface densities determined within r$_\mathrm{eff}$.}
\end{deluxetable*}

%%%%%%%%%%%%%%%%%%%%%%%%%%%%%%%%%%%%%%%%%%%%%%%%%%%%%%%%%%%%%%%%%%%%%%%%%%%%%%%%%%%%%
%%%%%%%%%%%%%%%%%%%%%%%%%%%%%%%%%% SPT0346-52 %%%%%%%%%%%%%%%%%%%%%%%%%%%%%%%%%%%%%%%
%%%%%%%%%%%%%%%%%%%%%%%%%%%%%%%%%%%%%%%%%%%%%%%%%%%%%%%%%%%%%%%%%%%%%%%%%%%%%%%%%%%%%
\subsection{SPT0346-52} \label{res0346}

%-----------------------------------------------------------------------------------
%%%%%%%%%%%%%%%%%%%%%%%%%%% FIGURE 3: SPT0346-52 SOURCE %%%%%%%%%%%%%%%%%%%%%%%%%%%%%
%-----------------------------------------------------------------------------------
\begin{figure}[htb]%
\includegraphics[width=\columnwidth]{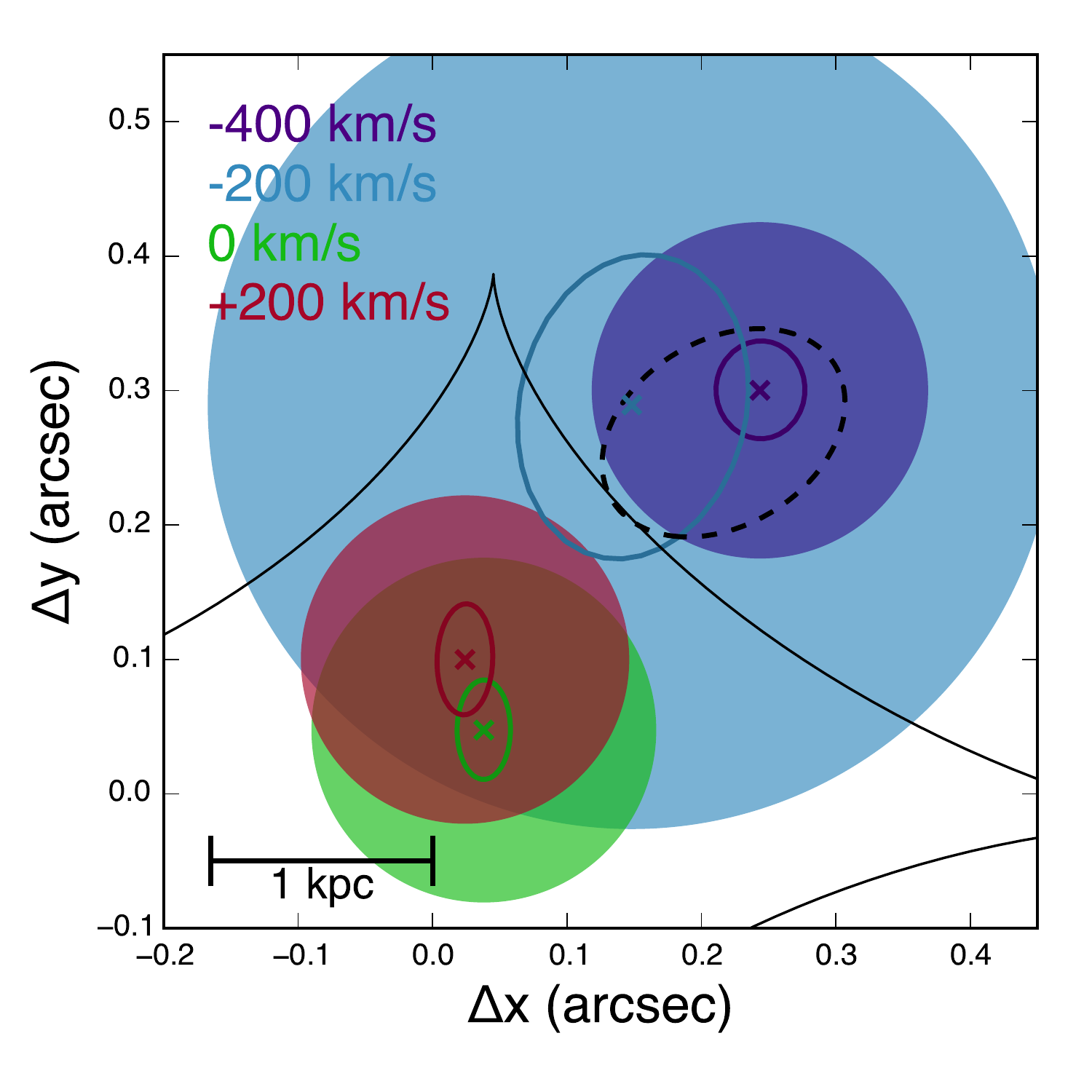}%
\caption{
Source-plane reconstruction of SPT0346-52.  Each CO(2--1) velocity channel modeled
is represented by a colored disk, truncated at the half-light radius.
Colored ``$\times$'' symbols and ellipses show the centroid and 1$\sigma$ positional
uncertainty on the centroid for each channel.  The FWHM and location of the 870\,\um
dust emission is shown as a thick dashed line, while the lensing caustic is shown
as a thin black line.
}%
\label{fig:src0346}%
\end{figure}
%-----------------------------------------------------------------------------------

As shown in Fig.~\ref{fig:images0346}, we detect the CO(2--1) emission from SPT0346-52
at peak significance of $>$$5\sigma$ in four consecutive 200\,\kms wide channels.  
A single symmetric Gaussian source-plane component for each
frequency channel is sufficient to model the observed
emission with residuals consistent with noise in all cases.  The derived 
source-plane structure is shown in 
Fig.~\ref{fig:src0346}, where we have truncated the models of each channel at the
half-light radius for clarity.  The line emission blueward of $\sim-100$\,\kms is
significantly offset from the redder emission.  We have modeled alternative
channelizations of the data, with velocity bins ranging from 100--400\,\kms, and all
channel widths point towards the same overall structure.  Derived source properties
are given in Table~\ref{tab:sourceprops}.

The structure seen in Fig.~\ref{fig:src0346} is difficult to interpret. At the depth
of these data, the velocity structure is not obviously consistent but not clearly
inconsistent with large-scale disc rotation or other bulk motion.  As we have only
modeled four consecutive velocity channels, it is difficult to rule out either
ordered or disordered kinematics. Taking the best-fit centroid of each velocity
component at face value, the molecular gas in SPT0346-52 appears more consistent
with a merging system than with the massive rotating discs seen in normal
star-forming galaxies at moderate redshift
by, e.g., \citet{forsterschreiber09} and \citet{tacconi13}. If, on the other hand,
the centroid of the -200\,\kms component is, in fact, between the bluer
and redder channels, a position-velocity curve resembling a rotating disc could
result.  Similar position-velocity diagrams were found by \citet{riechers08} and
\citet{deane13b} in the source-plane structure of lensed quasars at $z=4.1$ and
$z=2.3$, using data of similar significance to that presented here. These authors
interpreted their data as suggestive of rotation.  Deeper observations are
necessary to draw stronger conclusions about the velocity structure of SPT0346-52,
as this would allow lens modeling of narrower velocity bins.

The half-light radius of the background source is larger in low-J CO emission than in 
rest-frame 130\,\um continuum
emission, and the 130\,\um emission appears closely associated
with the blue half of the CO(2--1) emission.  This tentatively suggests that the 
star formation in SPT0346-52 is proceeding in a compact region embedded within a 
larger reservoir of molecular gas.
Alternatively, the CO(2--1) emission may trace a larger region of the galaxy
owing to its higher optical depth.
To test the degree to which a source with the
same size and location as either of the CO-emitting components at 0 and +200\,km/s could
contribute to the 130\,\um continuum emission, we re-fit the ALMA data with two 
source-plane components.  We fix the size and position of each source, leaving 
only the flux density of each source
as a free parameter.  The size and position of one source is fixed to that derived
from the ALMA data, while the other is fixed to the size and position of either of
the two red CO channels.  This is effectively a null test to determine how much flux
density
at rest 130\,\um could be emitted from the same region as the 0 or +200\,\kms CO emission.
This test indicates that a source co-located with either of the two reddest modeled
CO channels contributes $<3$\% (1$\sigma$ upper limit) of the total unlensed flux 
at 130\,\um.  This is also the fraction of \lir and, by proxy, SFR, that could arise
from these locations under the assumptions of a uniform dust temperature across
the source and no contribution from dust heating due to AGN activity.  
This limit is nevertheless consistent with CO/\lir ratios seen
in  local ULIRGs and $z>0.4$ main-sequence galaxies (e.g., 
\citealt{ivison11}; Aravena \etal \textit{in prep.}).

We note that our finding of a large CO spatial extent compared to dust
continuum is unlikely to be explained by the effects of interferometric filtering.
While only a limited range of spatial frequencies are probed by both the ATCA
and ALMA data, the data probe a similar range of radii in the $uv$ plane.
Additionally, as we fit directly to the visibilities, our model natively
reproduces this filtering. The models presented here recover 85--105\% of the 
flux observed in the most compact array configuration data. These data are included
in the modeling, so it is unsurprising that the total flux should be recovered well.

%%%%%%%%%%%%%%%%%%%%%%%%%%%%%%%%%%%%%%%%%%%%%%%%%%%%%%%%%%%%%%%%%%%%%%%%%%%%%%%%%%%%%
%%%%%%%%%%%%%%%%%%%%%%%%%%%%%%%%%% SPT0538-50 %%%%%%%%%%%%%%%%%%%%%%%%%%%%%%%%%%%%%%%
%%%%%%%%%%%%%%%%%%%%%%%%%%%%%%%%%%%%%%%%%%%%%%%%%%%%%%%%%%%%%%%%%%%%%%%%%%%%%%%%%%%%%
\subsection{SPT0538-50} \label{res0538}

%-----------------------------------------------------------------------------------
%%%%%%%%%%%%%%%%%%%%%%%%%%% FIGURE 4: SPT0538-50 SOURCE %%%%%%%%%%%%%%%%%%%%%%%%%%%%%
%-----------------------------------------------------------------------------------
\begin{figure}[htb]%
\includegraphics[width=\columnwidth]{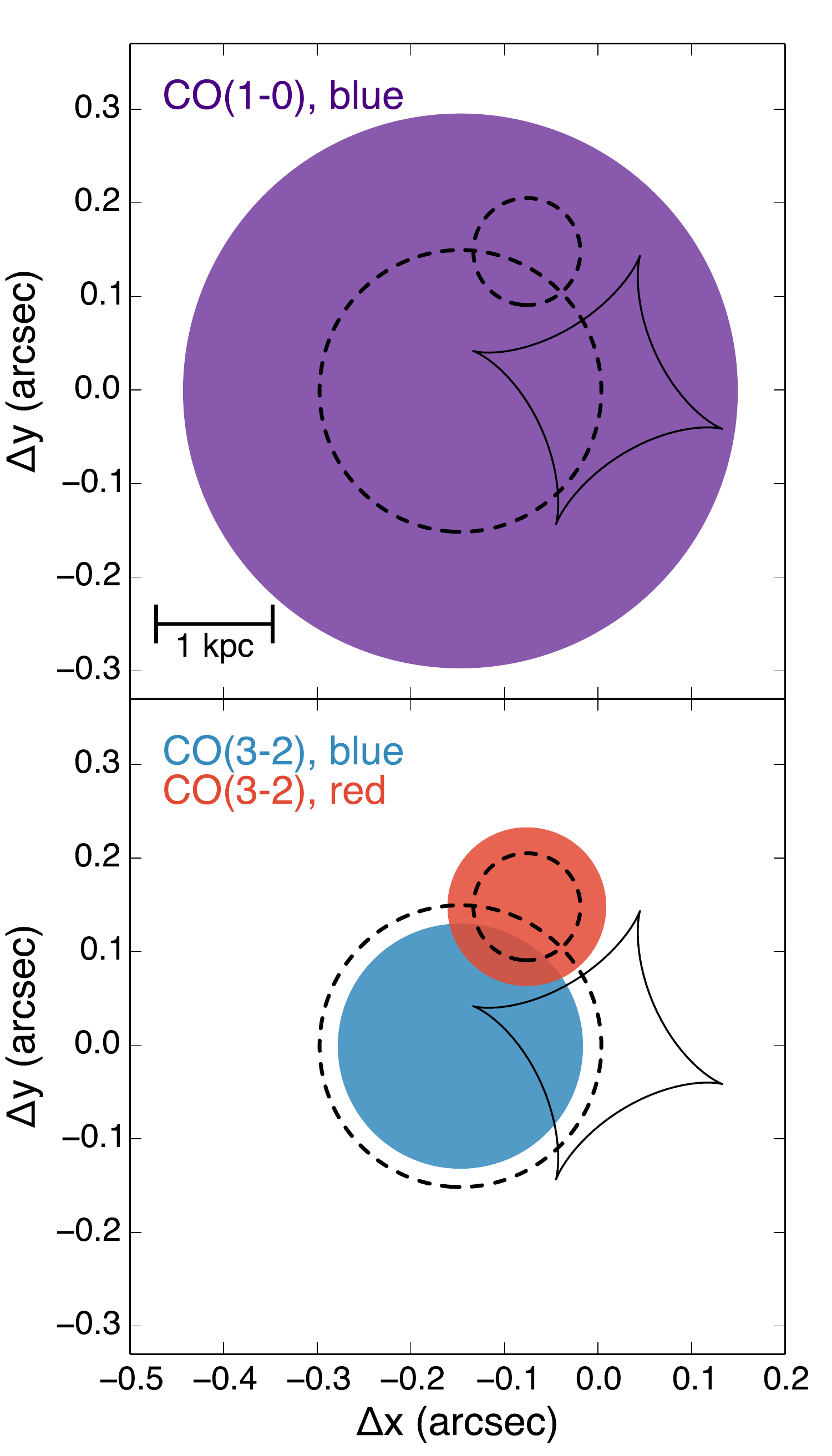}%
\caption{
Source-plane reconstruction of SPT0538-50.  As before, each component is truncated
at its FWHM. In both panels, the bright/compact and faint/extended rest-frame 230\,\um dust
components are represented by small and large dashed circles.  The best-fit
extent of the blue CO(1--0) velocity component is shown in the top panel, while
the bottom shows the extent of both the blue and red CO(3--2) velocity components.
Note that we are unable to model the red CO(1--0) velocity component, and that the
centroids of each CO component were fixed to the best-fit positions derived from
the ALMA continuum data.
}%
\label{fig:src0538}%
\end{figure}
%-----------------------------------------------------------------------------------

The integrated spectrum of this source shows two velocity components in CO(1--0)
emission separated by $\sim 350$\,km/s (Fig.~\ref{fig:images0538}; 
\citealt{aravena13,bothwell13b}), and so we model the emission
from each velocity component separately.  Unfortunately, our high-resolution CO(1--0) 
maps are too
shallow to detect the faint red velocity component.  However, this source was also
observed in CO(3--2) emission, marginally resolving the source and significantly
detecting both velocity components.  These observations also detected 
the observed-frame
3.3\,mm continuum emission at $\gtrsim 5 \sigma$.  Here, we discuss the results of the
lensing inversion of these three datasets in turn.  As before, we fix the parameters
of the lens to those derived from the ALMA 870\,\um data.

\subsubsection{Dust Continuum Emission}

As previously described and as found by \citet{hezaveh13}, we require two source-plane
components to fit the 870\,\um continuum emission observed by ALMA (rest-frame 230\,\um).
The source consists of
a faint, diffuse dust component of intrinsic flux density 1.4$\pm$0.4\,mJy, 
half-light radius $1.2\pm0.3$\,kpc
magnified by $\mu = 24\pm4.5$, and a brighter, more compact component of flux density 
$3.5\pm0.7$\,mJy and 
half-light radius $460\pm90$\,pc magnified by $\mu = 20\pm4$.

We have now also
detected dust continuum emission at observed-frame 3.3\,mm (rest-frame 870\,\um) in 
the line-free channels
of our ATCA 3\,mm data.  As the continuum is much more weakly detected in the ATCA data
than the ALMA data (Fig.~\ref{fig:images0538}), and as dust gives rise to the observed
emission at both wavelengths, we fit the ATCA data with two source-plane components
with positions and sizes fixed to the best-fit values derived from the ALMA data. This
leaves only the fluxes of each component as free parameters.  

Assuming the source-plane morphology of the dust emission is
the same at both rest-frame 230 and 870\,\um, we derive 
intrinsic fluxes of $55 \pm 15$\,\uJy for the bright, compact (A) component and
$5 \pm 9$\,\uJy for the faint, diffuse (B) component.
By calculating the flux ratio of the components at each MCMC step, we
rule out the possibility that the two components have the same 2.5:1 flux ratio
seen at 870\,\um at the $\sim 2.2\sigma$ level, tentative evidence that the two components
have different metallicities, dust temperatures and/or opacities.  
Further multi-wavelength 
high-resolution continuum observations could resolve this issue, though the atmosphere
limits such prospects at wavelengths shorter than observed-frame 350\,\um 
(rest-frame 92\,\um).

\subsubsection{CO(3--2) Emission}

In star-forming galaxies, the CO(3--2) transition is significantly
more luminous
than lower-lying transitions; for thermalized level populations, the 
integrated line flux scales
as $J^2$.  This, combined with the larger beam size of our ATCA 3\,mm
observations, allowed us to detect the fainter red velocity component in CO(3--2)
emission while this component remained undetected in CO(1--0).

As clearly seen in the images of Fig.~\ref{fig:images0538}, the 
spatial distributions of the blue and red CO(3--2) emission appear significantly
different. The red velocity component
appears similar to the rest-frame 870\,\um continuum 
emission simultaneously observed by ATCA.  This emission is
dominated by the bright, compact (A) dust component, as described in the previous section.
We begin exploring the lensing inversion
by fitting a single Gaussian profile to each velocity component, allowing the position,
size, and flux of each source to vary.  The best-fit position of the blue velocity
component is within the 1$\sigma$ uncertainties of the location of the diffuse B
dust component seen in the ALMA data, and inconsistent with the location of the
compact A component.  While the positional uncertainties are large
($\sim$0.13"), the best-fit position of the red velocity component is consistent with
the location of the bright, compact dust component seen in the ALMA maps.  This suggests
that the two velocity components are in fact associated with the two dust components.

Motivated by this association, we re-fit the CO(3--2) data of both the red and blue
velocity components with \textit{two} source-plane components each, fixing the
positions of each component to the best-fit positions of the dust components in
the continuum lens model, allowing the source size to remain a free parameter. 
This is a test to determine whether the CO emission
of each velocity component can be uniquely associated with one of the two
dust components.  The results of this test strengthen the hypothesis that each
velocity component is associated with only one of the two dust components --
for each velocity component, the lens modeling prefers that only one of the
two source-plane components have positive flux and non-infinite size.
Again, we find that the brighter blue velocity component is associated with the faint,
diffuse (B) dust component, and the fainter red velocity component is associated with
the bright, compact (A) dust component.

Figure~\ref{fig:src0538} shows the results of modeling the CO(3--2) emission, where
we have fixed the locations of the blue and red velocity components to the locations
of the faint and bright dust components, respectively.  For each velocity channel,
we allow the flux and size of the modeled source-plane component to vary.  The models
imply that the CO(3--2)-emitting molecular gas has approximately the same extent
as the dust emission in both components.   

Having associated the two dust components seen in the lensing reconstruction with
the two separate velocity components seen in the integrated CO line spectra, we
strengthen the argument that SPT0538-50 is indeed a pair of merging galaxies, as also
posited by \citet{bothwell13b}.  Their arguments, based on the high SFR surface density,
high specific SFR, and suppressed fine structure lines of SPT0538-50, 
are confirmed based on our high-resolution
kinematic observations.  The two merging galaxies are separated by 1.3\,kpc in projection
and $\sim375$\,\kms in velocity.  This is comparable to what is seen in the local
ULIRG Arp220, a late-stage merger which shows two nuclei separated by approximately
400\,pc and $\sim250$\,\kms \citep[e.g.,][]{scoville97,sakamoto09}.  
In contrast, SPT0538-50 appears to be in a more compact merger than the $z=5.24$ 
\textit{Herschel}-selected lensed galaxy HLS0918 studied by \citet{rawle14}, 
which consists of four spectral components separated by 4\,kpc and $\sim840$\,\kms.
SPT0538-50 appears to conform with the idea that most DSFGs reach their extreme
SFRs through major merger activity \citep[e.g.,][]{engel10}.

\subsection{CO(1--0) Emission}

As seen in Fig.~\ref{fig:images0538}, our observations of CO(1--0) are insufficiently
deep to detect the faint, red velocity component in our high-resolution imaging data.
We do, however, clearly detect the brighter blue
velocity component at comparable resolution to the ALMA 870\,\um data.  The morphology
of the CO(1--0) emission is clearly different from that of the dust continuum emission,
which is dominated by the bright, compact dust component described previously.  While
we cannot model the red line component in CO(1--0), we proceed by discussing the blue
line component, which can be modeled.

Having established in the previous section that the blue line component is 
spatially associated with the faint, diffuse dust component, we model the 
blue CO(1--0) emission with a circularly-symmetric Gaussian source-plane 
component with position fixed to that derived from the ALMA continuum data.  
The free parameters are the source flux and size, as in our models of the
CO(3--2) emission.  The best-fit model source, shown in Fig.~\ref{fig:src0538} and
described in Table~\ref{tab:sourceprops}, leaves residuals consistent with 
noise.  Comparison of the CO(3--2) and CO(1--0) intrinsic line fluxes implies
a CO brightness temperature ratio of $r_{31} \sim 0.6$, similar to the ratios
determined for other high-redshift DSFGs and slightly lower than the average 
ratio for the SPT DSFGs themselves \citep[e.g.,][]{danielson11,bothwell13,spilker14}.
Similar to what
was seen in CO(2--1) in SPT0346-52, we find that the CO(1--0) emission in SPT0538-50
is significantly extended compared to the dust emission.  Again, this may indicate
that the star formation in this system is proceeding in a compact region embedded in
a larger reservoir of molecular gas.  We return to this discussion in the next section.

%%%%%%%%%%%%%%%%%%%%%%%%%%%%%%%%%%%%%%%%%%%%%%%%%%%%%%%%%%%%%%%%%%%%%%%%%%%%%%%%%%%%%
%%%%%%%%%%%%%%%%%%%%%%%%%%%%%%%%%%% Discussion %%%%%%%%%%%%%%%%%%%%%%%%%%%%%%%%%%%%%%
%%%%%%%%%%%%%%%%%%%%%%%%%%%%%%%%%%%%%%%%%%%%%%%%%%%%%%%%%%%%%%%%%%%%%%%%%%%%%%%%%%%%%
\section{Discussion} \label{discussion}

\subsection{Source Sizes and Differential Magnification} \label{diffmag}

As can be seen in Figures~\ref{fig:src0346} and \ref{fig:src0538}, we find that
the molecular gas reservoirs traced by low-J CO transitions have larger half-light radii
than the emission from the dust continuum.  If we attribute all of the dust emission
to star formation (as opposed to, for example, dust heated by AGN activity), this
implies that the intense star formation in these galaxies
is limited to relatively small regions embedded in much larger reservoirs of
molecular gas.  The kinematics of both galaxies are plausibly consistent with the
disruption of secular rotation, causing vast amounts of molecular gas to fall towards
dense, compact star-forming regions.

In SPT0538-50, we find that the CO(1--0) is similarly extended compared to
CO(3-2), by more than a factor of 2$\times$ in the blue velocity component. Size
differences of this level were reported by \citet{riechers11}, comparing the and CO(1--0) sizes of the lensed DSFG SMM\,J09431+4700. Larger physical
extents for low-J CO emission were also inferred by those authors and 
\citealt{ivison11} by comparing the CO line widths between low- and mid-J CO
transitions.
In contrast with the CO(1--0) emission, we find the CO(3--2) emission in SPT0538-50
to have roughly equal half-light radius as the star formation traced by the dust
continuum.  This can be taken as evidence that the CO(3--2)
emission is more directly associated with ongoing star formation in this galaxy, in
agreement with studies both locally \citep[e.g.,][]{wilson09} and at 
high-redshift \citep{bothwell10,tacconi13} which
find an approximately linear relationship between the CO(3--2) luminosity and SFR. 
Indeed, high gas excitation conditions ($\tkin \sim 50$\,K, 
$\nht \sim 1000$\,\percc) are needed 
to achieve brightness temperature ratios near unity.
A similarity in size between CO(3--2) and 
stellar light was also seen by \citet{tacconi13}, who used rest-frame $B$-band
\textit{HST} images to determine the extent of the star formation in a large sample
of $z \sim 1-2$ normal star-forming galaxies.  

\begin{figure}[htb]%
\includegraphics[width=\columnwidth]{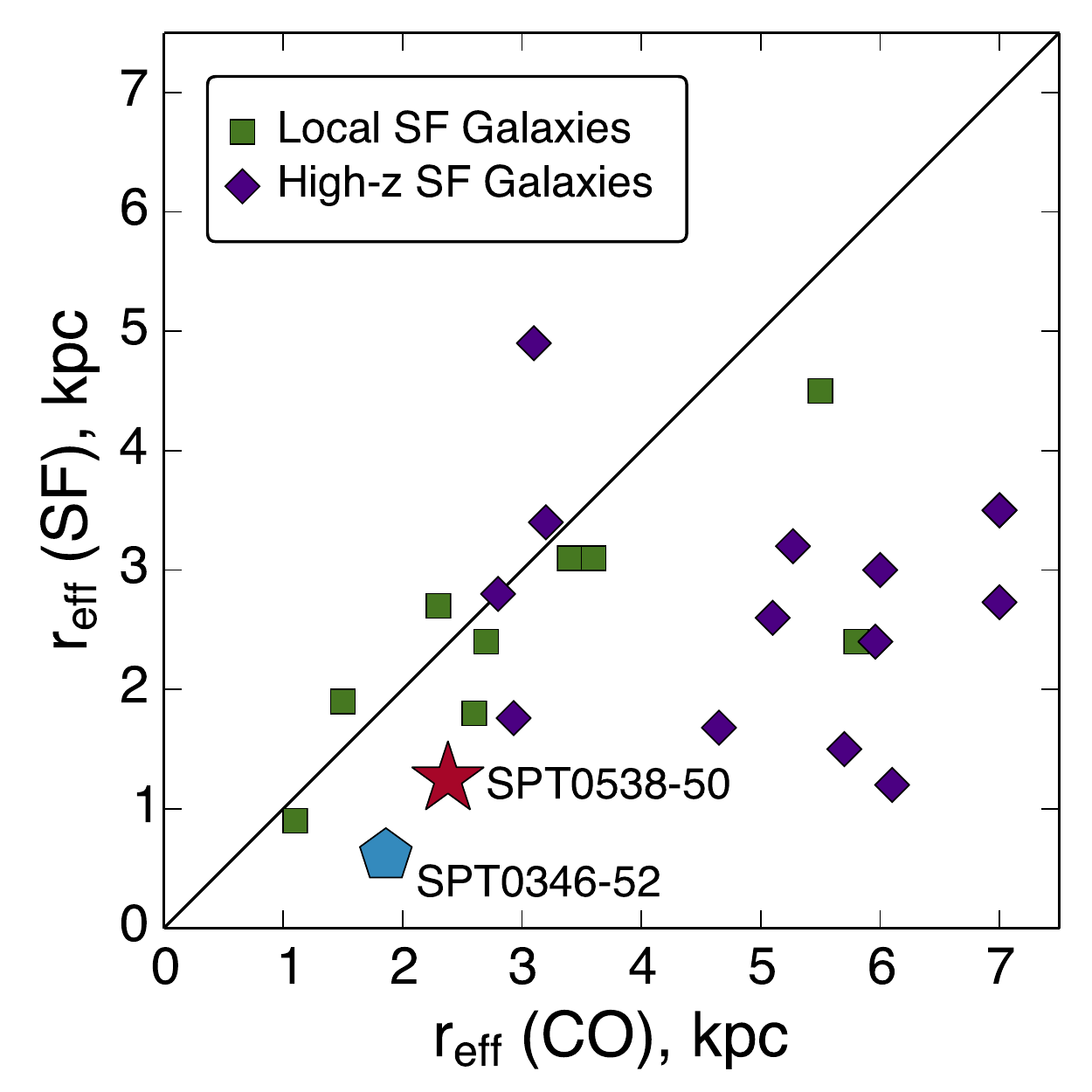}%
\caption{
Relative sizes of star formation and molecular gas in galaxies.  CO source sizes
are confined to those objects for which low-$J$ resolved measurements have been
made.  Star formation sizes are derived either from rest-frame UV measurements,
dust continuum emission, or both.  The solid line indicates 
r$_\mathrm{eff}$(SF) = r$_\mathrm{eff}$(CO). The high-redshift sample is drawn from 
\citet{younger08,daddi10,ivison11,fu12,hodge12,walter12,fu13,ivison13}. The local
galaxies are drawn from \citet{regan01} and \citet{leroy08}, where we have scaled
the CO sizes in Regan \etal to match the distances in Leroy \etal.
}%
\label{fig:sizes}%
\end{figure}

While measurements are few, in Fig.~\ref{fig:sizes} we plot the half-light radii
of star formation (traced by the rest-frame UV, dust continuum emission, or both)
and molecular gas reservoirs for local and high-redshift 
sources.  We restrict
this comparison to those galaxies with measured sizes in CO transitions with 
$J_\mathrm{up} < 3$, since the effective source size changes significantly
as a function of observed transition.  Physically large molecular gas reservoirs are
common, with an average star formation area filling factor of $\sim 55$\%, similar
to the difference in size between gas and SF in local galaxies seen by
\citet{bigiel12} and \citet{zahid14}.
This difference in size can be a potential source of bias when calculating
surface-density quantities using sizes derived at different wavelengths -- for example,
using the source sizes determined from a SF tracer to calculate
\siggas would lead to an over-estimation of the 
\textit{average} surface density by a factor of 80\%.

%-----------------------------------------------------------------------------------
%%%%%%%%%%%%%%%%%%%%%%%%%%% FIGURE 5: SOURCE SIZES %%%%%%%%%%%%%%%%%%%%%%%%%%%%%%%%%
%-----------------------------------------------------------------------------------
%-----------------------------------------------------------------------------------

This difference in source size and structure at different emitting wavelengths leads 
to differential magnification. 
Although gravitational lensing is
achromatic, galaxies appear morphologically different at different wavelengths, leading
to wavelength-dependent variations in the lensing magnification
\citep[e.g.,][]{blain99c,serjeant12,hezaveh12a}.  In both sources 
studied here, the large total extent of the CO emission relative to the dust continuum
emission and the proximity of the sources to the lensing caustics lead to differences 
in magnification between dust and molecular gas of up to $\sim50$\%.  In SPT0346-52,
our data indicate that the gas emitting at positive systemic velocities is magnified by
approximately a factor of 2$\times$ more than either the rest-frame 130\,\um continuum
or the gas at bluer systemic velocities.  Such a magnification gradient may
in fact be a common feature of observing lensed galaxies in spectral lines in which
galaxies are expected to have large physical extents. \citet{riechers08} 
and \citet{deane13b} both find
similar levels of differential magnification as a function of velocity in observations of
low-J CO emission in the lensed quasars PSS~J2322+1944 and IRAS~F10214+4724,
respectively.  These results indicate that differential magnification can cause
significant distortion of the global SEDs of lensed galaxies, reaffirming the need
to account for its effects when comparing lensed and unlensed sources.

\subsection{The CO-H$_2$ Conversion Factor} \label{alphaco}

With the wealth of high-resolution data on these two sources, we are presented with
the opportunity to constrain the conversion factor between CO luminosity and molecular
gas mass, \alphaco, through multiple means.  This allows us to measure the extent to
which different methods of determining \alphaco are consistent within a single system.
Below, we briefly describe each technique as applied to SPT0346-52 and SPT0538-50.  
In each case, we assume the mass in the form of atomic hydrogen gas is negligible,
as seems appropriate for dense, highly star-forming systems.
We also assume the CO(2--1) line observed in SPT0346-52 is thermalized and
optically thick, with $L_{\rm{CO(1-0)}}' = L'_{\rm{CO(2-1)}}$. Given the vigorous
star formation proceeding in this warm, dense system and that 
T$_\mathrm{CMB}(z=5.7) = 18$\,K $>$ E$_\mathrm{up,CO(2-1)}/k_B$, 
this assumption is justified.  This ratio is also justified both 
observationally (see, e.g., Fig.~45 of \citealt{casey14}) and theoretically 
\citep{narayanan14}. The results of this section are summarized in
Fig.~\ref{fig:alphaco}.

\subsubsection{Gas-to-Dust Ratio}

Gas and dust are widely observed to be well-mixed in galaxies, with dust comprising
approximately 1\% of the mass in the ISM \citep[e.g.,][]{sandstrom13,draine14}. Thus,
if the mass in dust ($M_d$) and the gas-to-dust ratio (\gdr) can be estimated, \alphaco
can be determined simply as $M_d \gdr / \lprime$.

Extensive \textit{Spitzer} and \textit{Herschel} observations of the 
dust continuum emission in galaxies
have rapidly advanced our understanding of \gdr and the nature of the dust emission.
The metallicity of the ISM affects
\gdr, so that low-metallicity systems have more gas per unit dust mass.  Meanwhile,
most far-infrared dust SEDs can be accurately modeled using only a small number of
free parameters.  One common method is to fit the dust SED with a single-temperature
modified blackbody function. However, such single-temperature models generally
underestimate the dust mass in galaxies by a factor of $\sim2\times$
\citep{dunne01,dale12}. Dust at a 
single temperature cannot simultaneously fit both the long- and short-wavelength 
sides of the SED, which has contributions from dust heated to a range of 
equilibrium temperatures.  This effect has been explored by \citet{dunne00} and
\citet{dale12}, and indicates that a more sophisticated, multi-component approach
is needed.  Such approaches have been developed by, e.g., \citet{draine07} and
\citet{dacunha08}.  The theoretical models of \citet{draine07} assume the dust
is exposed to a power-law distribution of starlight intensities, while the MAGPHYS
code of \citet{dacunha08} decomposes the dust emission into different 
physically-motivated temperature regimes.  The \citet{draine07} models have been
used to constrain \alphaco in samples of normal and intensely star-forming
$z \sim 0.5-4$ \textit{Herschel}-selected star-forming galaxies observed 
by \citet{magdis12} and \citet{magnelli12}.

Calculation of the dust mass requires knowledge of the dust mass absorption
coefficient, defined here as \dustkappa, the value of the coefficient at
rest-wavelength 870\,\um, in units of m$^2$\,kg$^{-1}$. The dust mass scales as 
$\kappa_{\rm{870\,\um}}^{-1}$.  Most estimates of \dustkappa for 
dust in nearby, Milky Way-like
galaxies are close to either 0.045 
\citep{li01,draine07,scoville14} or 0.075 \citep{ossenkopf94,dunne00,james02}. The
choice of dust opacity, then, leads to a difference in dust mass of $\sim$70\%.
In this work, we adopt the dust models implemented in MAGPHYS \citep{dacunha08},
which uses the dust opacity coefficient of \citet{dunne00}, \dustkappa $= 0.077$.
This value for the mass absorption coefficient is commonly used for other 
high-redshift rapidly star-forming systems.

Finally, to calculate the total gas mass from the dust mass, we require
knowledge of the gas-to-dust ratio, \gdr, which is known to vary with metallicity.
As we have essentially no constraints on the metallicities
of these systems, we adopt the average \gdr determined by \citet{sandstrom13} in a
large sample of local star-forming galaxies with approximately solar
metallicities, $\gdr = 72$ with $\sim$0.2\,dex scatter.  We note that the 
derived \alphaco conversion factors are
linearly proportional to the assumed \gdr, and \gdr itself likely varies 
approximately linearly with metallicity \citep{leroy11}. 

We determine the dust masses of SPT0346-52 and SPT0538-50 by fitting to the photometry
given in \citet{weiss13} and \citet{bothwell13b}, respectively.  In both cases, we 
assume that differential
magnification of the dust emission is insignificant, and for SPT0538-50, we assume
the flux at each wavelength is divided between the two components in the same 
ratio as at rest-frame 230\,\um. This
yields dust masses of $2.1\pm0.3 \times 10^9$\,\msol for SPT0346-52 and $9.0\pm1.3$
and $3.7 \pm 0.6 \times 10^8$\,\msol 
for the compact and diffuse components of SPT0538-50, where the uncertainties are
statistical only.  For SPT0346-52, this dust 
mass and the source size derived from the ALMA lens model imply that the dust reaches unit
optical depth by rest-frame $\sim$300\,\um, longer than the canonical wavelength
of $\sim$100\,\um.  Lower optical depths are possible by raising the dust temperature
or effective source size. The prevalence of this effect in samples of lensed DSFGs
will be explored in more detail by Spilker et~al., \textit{in prep.}

Using our adopted \gdr and dividing by the sum of the intrinsic CO luminosities 
of each channel yields a measurement of $\alphaco = 2.2\pm0.6$ for
SPT0346-52.  For SPT0538-50, we calculate \alphaco for the blue velocity component
only, yielding $\alphaco = 1.7\pm0.4$.  Note that the uncertainties in these calculations
account only for the statistical errors in the dust SED fitting and lens modeling
procedure, and neglect systematic uncertainties in \dustkappa and \gdr, which are
of order 100\%.

\subsubsection{Dynamical Constraints}

The CO-H$_2$ conversion factor can also be constrained using estimates of the total
dynamical mass of galaxies and a process of elimination -- the molecular gas mass
is the remainder after all other contributions to the dynamical mass have been
subtracted (e.g., stars, dark matter, HI gas, dust, etc.).  Assuming that stars, molecular
gas, and dark matter make up the vast majority of the total mass, then, we have
\begin{equation} \label{eq:mdyn}
M_\mathrm{dyn} = M_* + \alphaco\lprime + M_\mathrm{DM},
\end{equation}
where each of these quantities is measured within the same effective radius 
defined by the extent of the CO emission.

Measuring each of these quantities is fraught with assumptions and systematic
uncertainties. The stellar mass of SPT0538-50 has been estimated by \citet{bothwell13b}
as $3.3 \pm 1.5 \times 10^{10}$\,\msol, while Ma et~al., \textit{in prep.} only
place an upper limit on the stellar mass of SPT0346-52, based on
SED fitting to optical through far-infrared photometry. 
These estimates assume the stellar light is
magnified by the same factor as the dust emission, which may be inaccurate by up to a
factor of $\sim$50\% (see Section \ref{diffmag} above and \citealt{calanog14}). 
The dark matter content within the region traced by CO
is highly uncertain, but we adopt a contribution of 25\% for consistency with the 
literature \citep{daddi10}, based on observations of $z \sim 1.5 - 2$ disk galaxies.

A crude estimate of the dynamical mass can be derived from
\begin{equation} \label{eq:mdyn}
M_\mathrm{dyn} = \gamma R \Delta V^2 / G,
\end{equation}
with $\Delta V$ the FWHM line width, $R$ the source effective radius, and
the gravitational constant $G$. Here the pre-factor $\gamma$ accounts for the detailed
geometry of the source, and is of order unity.  As the geometry of the two objects
studied here is only somewhat constrained, we adopt $\gamma = 1$ for simplicity.
Literature values of $\gamma$ range from $\gamma \sim 0.3$ 
\citep[e.g.,][]{neri03,tacconi08,daddi10} 
to $\gamma = 1.2$ \citep[e.g.,][]{bothwell13}.  In the following, we quote only
statistical uncertainties on the derived values of \alphaco, ignoring the much
larger systematic uncertainties on the stellar mass, dark matter fraction, and
true source geometry.

For SPT0538-50, using the spatial separation of the two components of $1.35 \pm 0.16$\,
kpc and a line width of 490\,\kms (from the full line profile), the above equation
yields $M_\mathrm{dyn} \sim 7.5 \pm 0.9 \times 10^{10}$\,\msol. Combined with the stellar
mass estimate given above and a 25\% dark matter fraction, this yields
\alphaco $= 1.4 \pm 0.8$.  If the stellar mass is entirely concentrated in the blue
velocity component, the implied gas fraction of the blue component is 
$f_\mathrm{gas} = M_\mathrm{gas}/(M_\mathrm{gas}+M_\mathrm{star}) \sim 40$\%. As
the red component must also contain some gas, 40\% is a lower limit to
the gas fraction of the entire system.

We estimate the dynamical mass of SPT0346-52 using the separation in space
and velocity of the emission at -400\,\kms and +200\,\kms. This yields
$R = 1.8 \pm 0.2$\,kpc and $M_\mathrm{dyn} = 1.5\pm0.2 \times 10^{11}$\,\msol.  
The stellar mass upper limit calculated by Ma et~al., \textit{in prep.} is
a factor of several higher than the dynamical mass we have estimated. To constrain
\alphaco, we instead assume a range of gas fractions of
$f_\mathrm{gas} = 0.3-0.8$. This range has been observed by \citet{tacconi13}
at lower redshifts ($z \sim 1-2$), and appears to evolve slowly with redshift
at $z>2$ (e.g., \citealt{bothwell13}; Aravena et~al., \textit{in prep.}).  This range in
gas fraction leads to a range of $\alphaco = 0.5 - 1.3$; we adopt
$\alphaco = 0.9 \pm 0.5$, where the uncertainty reflects only the range of 
$f_\mathrm{gas}$ and the statistical uncertainty on $M_\mathrm{dyn}$.

Given the large systematic uncertainties inherent in each step of these calculations, 
we estimate that the uncertainty in these derived conversion factors is at least
a factor of 2$\times$.  Improving these estimates would require extensive high
resolution multi-wavelength observations in order to better constrain the dynamical
and stellar masses.  Refined estimates of the dark matter contribution would be even
more challenging, and the best option may simply be to use the dark matter content
derived from hydrodynamical simulations of galaxies.

\subsubsection{CO Luminosity Surface Density}

A third estimator of the CO-H$_2$ conversion factor was developed by
\citet{narayanan12}, based on hydrodynamical simulations coupled with dust and
line radiative transfer.  Those authors developed a suite of simulated 
galaxies in isolated and merging systems and provided a fitting formula to \alphaco
that depends solely on the CO line intensity and metallicity:
\begin{equation} \label{alphasigmaco}
\alphaco = 10.7 \times \Sigma_\mathrm{CO}^{-0.32} / (Z/Z_\odot)^{0.65},
\end{equation}
where $\Sigma_\mathrm{CO}$ is the CO luminosity surface density in units of K\,km/s.
As in the dust-to-gas ratio method, we again assume solar metallicity for both DSFGs
considered here; lowering the metallicity to half solar would increase these estimates
by approximately 50\%.
For SPT0346-52, we average the values of \alphaco determined for each modeled
channel to find a value of $\alphaco = 0.78\pm0.11$. As we could not model the red velocity
component in SPT0538-50, for this galaxy we calculate \alphaco
for the blue velocity component only, yielding $\alphaco = 1.4 \pm 0.2$.

\subsubsection{Summary of CO--H$_2$ Conversion Factor Measurements}

We have constrained the CO--H$_2$ conversion factor in the two objects presented
here using three independent methods.  The three methods show reasonably good 
agreement with each other, with average values of $\alphaco = 1.3$ for
SPT0346-52 and $\alphaco = 1.5$ for SPT0538-50.  The uncertainty in these 
estimates is dominated by systematic, rather than statistical, errors of
$\sim$100\% for each method.  Given the large systematic
uncertainties inherent to each method, the general agreement between the three
techniques is encouraging. Resolved CO and dust
continuum observations of a larger sample of objects could reveal systematic
differences between the various methods.

\begin{figure}[htb]%
\includegraphics[width=\columnwidth]{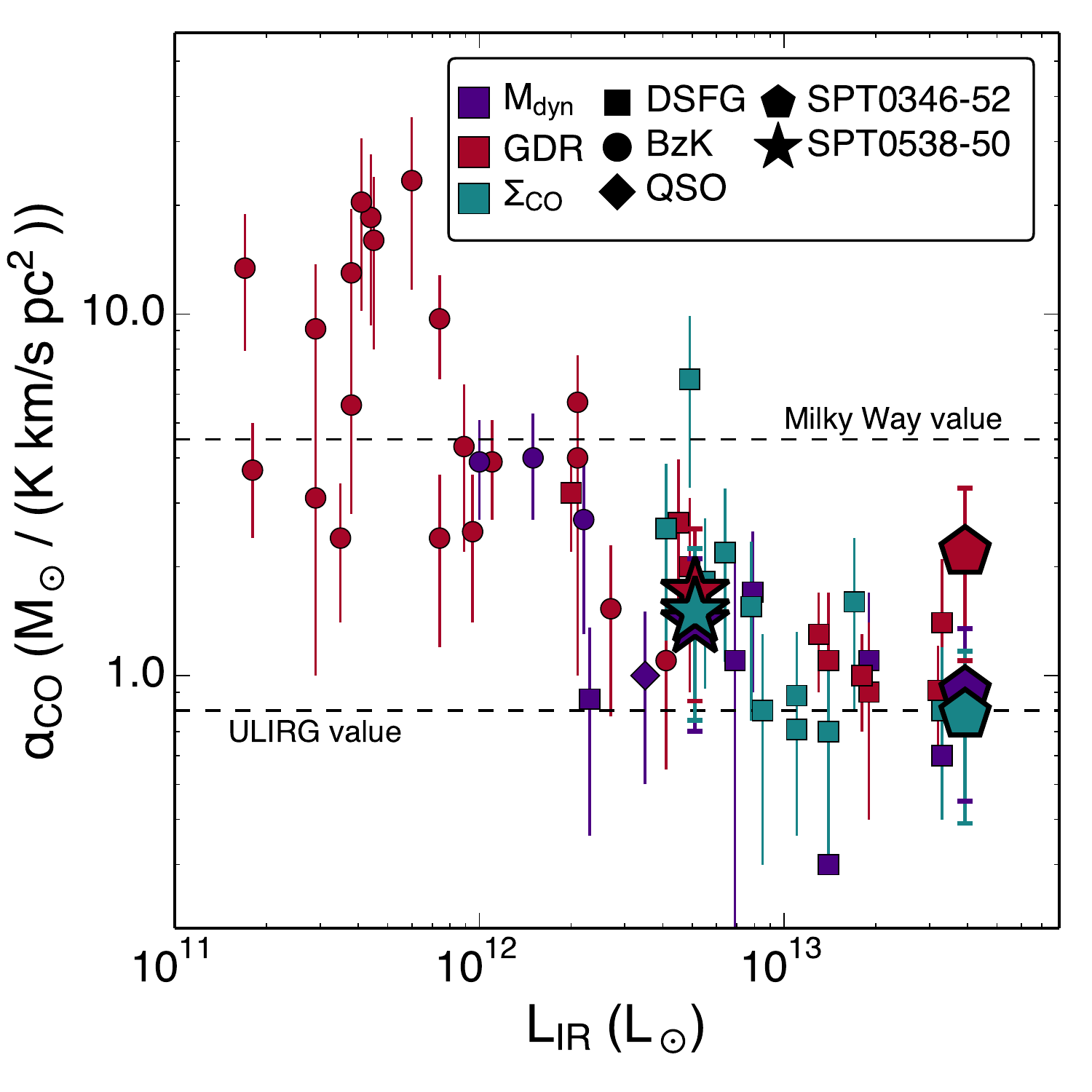}%
\caption{
Summary of constraints on the CO-H$_2$ conversion factor at $z>1$ from the
literature. Objects are
color-coded by the method used to constrain \alphaco and shape-coded by type of
object.  SPT0346-52 and SPT0538-50 both appear to follow the general trend of
decreasing \alphaco with increasing \lir.  Where applicable, \lir has been
corrected for lensing magnification. If not given in the original references,
measurement uncertainties for the literature objects are shown as 50\%. Objects with
constraints from multiple methods are shown multiple times at the same \lir.
Literature objects
are compiled from \citet{daddi10,ivison11,magdis11,swinbank11,fu12,hodge12,
magdis12,magnelli12,walter12,deane13b,fu13,hodge13,ivison13,messias14}.
}%
\label{fig:alphaco}%
\end{figure}

For both objects,
the derived conversion factors are similar to those determined for other
rapidly star-forming objects, exhibiting low values of \alphaco similar
to most DSFGs and unlike more quiescently star-forming objects.
In Figure~\ref{fig:alphaco}, we place our measurements
in the context of other studies that have also constrained the conversion factor 
in a wide variety of galaxies at
$z > 1$.  No clear bimodality between ``Milky Way-like'' and ``ULIRG-like'' values 
is seen, as would be expected from a heterogeneous collection of galaxies with
varying ISM properties.  Indeed, the sample used to determine the  canonical 
``ULIRG-like'' value of \alphaco by \citet{downes98} also showed a fair amount of
variation.
This lack of bimodality (as noted by \citealt{narayanan12}
and discussed in the next section) can influence the form of the Schmidt-Kennicutt
star formation relation.

%-----------------------------------------------------------------------------------
%%%%%%%%%%%%%%%%%%%%%%%%%%% FIGURE 6: ALPHA_CO SUMMARY %%%%%%%%%%%%%%%%%%%%%%%%%%%%%
%-----------------------------------------------------------------------------------
%-----------------------------------------------------------------------------------

\subsection{The Star Formation Relation at $z > 2.5$} \label{KSrel}

\begin{figure}[htb]%
\includegraphics[width=\columnwidth]{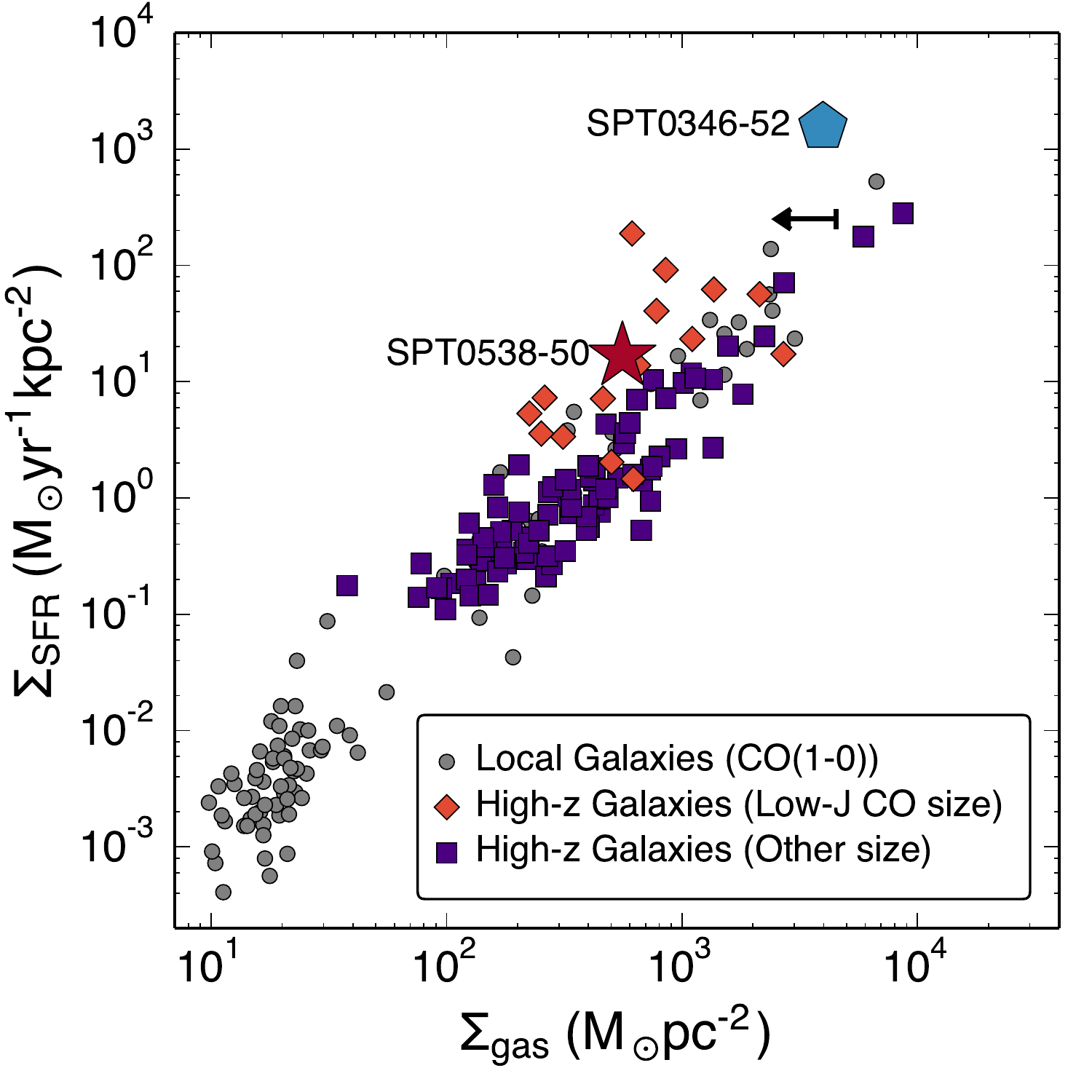}%
\caption{
The Schmidt-Kennicutt star formation relation.  For all galaxies, we use CO-H$_2$
conversion factors using the \citet{narayanan12} formula.  The high-redshift 
galaxies are separated into those that have been resolved in a low-$J$ CO
transition and those for which a higher line was observed (with a size determined
from other means, generally the rest-frame optical/UV size).  The black arrow shows
the effect of the average correction to \siggas that results from the difference
in size between the low-$J$ CO emission and the rest-frame UV or dust emission, 
as seen in Fig.~\ref{fig:sizes}. The local galaxy
sample is from \citet{kennicutt98b}, while the high-redshift sample is compiled 
from \citet{younger08,genzel10,daddi10,ivison11,magdis11,fu12,hodge12,
walter12,fu13,ivison13,tacconi13,hodge15}.
}%
\label{fig:sfrel}%
\end{figure}

We now turn our attention to the Schmidt-Kennicutt \sigsfr--\siggas relation.
Figure~\ref{fig:sfrel} presents samples of low- and high-redshift galaxies for comparison 
to the targets of this work. Our lens modeling and FIR SED provide measurements of 
the total star formation rate, molecular gas mass (inferred from low-$J$ CO measurements 
and the \alphaco measurements of the previous section), and sizes of the 
star-forming and molecular gas regions. The \sigsfr and \siggas values for these 
two objects show them to lie at the upper edge of the distribution of galaxies in 
this plane. 

To interpret the offset between the properties of the SPT galaxies and the larger sample
consisting of both normal and more highly star-forming galaxies, 
we must consider the measurements that are required to construct such a plot. 
The high-redshift comparison galaxies in Fig.~\ref{fig:sfrel} are divided into two samples 
depending on whether their molecular gas sizes are determined from low-$J$ CO 
(orange diamonds) 
or from other measurements, including star formation tracers and higher-$J$ CO 
(purple squares). The galaxies 
with low-$J$ CO size measurements, which provide determinations of \siggas 
that are most similar 
to those made at low redshift, are more consistent with the SPT sources than 
the other high-redshift subsample. 
The typical procedure to convert from higher-$J$ CO 
measurements to \siggas involves correcting the CO luminosity for 
subthermal excitation to arrive at 
the CO(1--0) equivalent, using \alphaco to infer the molecular gas mass, 
and division by one of 
the available size measurements to get the surface density. 
However, in Section~\ref{diffmag} we found that other measurements 
of galaxy size systematically 
underestimated the low-$J$ CO size by a factor of $\sim$1.3, which corresponds to a 
1.7$\times$ underestimate in the area and overestimate in \siggas. This error is shown as a 
horizontal arrow in Fig.~\ref{fig:sfrel}, and is very similar to the \
offset seen between the two high-redshift 
galaxy samples. Clearly, care is needed when placing galaxies observed 
in heterogeneous ways on the SF relation.
Further comprehensive studies of low-$J$ CO emission
by the VLA or ATCA offer the potential to resolve this issue by directly
comparing the effective radii of high-redshift galaxies at various
CO transitions.

%-----------------------------------------------------------------------------------
%%%%%%%%%%%%%%%%%%%%%%%%%%% FIGURE 7: SF-GAS RELATION %%%%%%%%%%%%%%%%%%%%%%%%%%%%%%
%-----------------------------------------------------------------------------------
%-----------------------------------------------------------------------------------

\section{Conclusions} \label{conclusions}

We have presented spatially and spectrally resolved images of two 
gravitationally lensed dusty
star-forming galaxies at redshifts $z=2.78$ and $z=5.66$.  
SPT0346-52 is among the most intrinsically luminous DSFGs, while the luminosity of SPT0538-50 is
more typical of the DSFG population.
Using a visibility-based
lens modeling procedure, we have shown that SPT0346-52 has complex
dynamics, and confirmed the merger hypothesis in SPT0538-50.  By comparing with
lens models derived from ALMA observations of the dust continuum in each galaxy,
we find that the difference in magnification between the molecular gas and dust
varies between 0 and 50\%, mostly due to the larger physical extent of the gas
compared to the area of active star formation.  In SPT0538-50, we have shown 
that the physical extent
of the CO emission decreases with increasing transition, with the CO(3--2) 
emission being roughly the same size as the dust continuum.  We have constrained 
the CO--H$_2$ conversion factor
via three independent methods, finding values near those expected for highly
star-forming systems.  The three methods agree reasonably well when applied to
these two objects; further in-depth studies may be able to discern systematic
differences between the various methods.  Finally, we have placed these two
objects on the Schmidt-Kennicutt star formation relation, finding that they lie
along the upper envelope of vigorously star-forming systems.  Part of this offset
may be explained by the different effective source sizes of the CO emission as a
function of observed transition, an effect that should be taken into account as larger 
samples of spatially-resolved high-redshift molecular gas measurements become available.

\acknowledgements
{
J.S.S., D.P.M., and J.D.V. acknowledge support from the U.S. National Science Foundation under grant No. AST-1312950 and through award SOSPA1-006 from the NRAO.
M.A. acknowledges partial support from FONDECYT through grant 1140099.
This material has made use of the El Gato high performance computer, supported by the U.S. National Science Foundation under grant No. 1228509.
The Australia Telescope Compact Array is part of the Australia Telescope National Facility, which is funded by the Commonwealth of Australia for operation as a National Facility managed by CSIRO.
This paper makes use of the following ALMA data: ADS/JAO.ALMA \#2011.0.00957.S and \#2011.0.00958.S.  ALMA is a partnership of ESO (representing its member states), NSF (USA) and NINS (Japan), together with NRC (Canada) and NSC and ASIAA (Taiwan), in cooperation with the Republic of Chile. The Joint ALMA Observatory is operated by ESO, AUI/NRAO and NAOJ.  The National Radio Astronomy Observatory is a facility of the National Science Foundation operated under cooperative agreement by Associated Universities, Inc. 
The SPT is supported by the National Science Foundation through grant PLR-1248097, with partial support through PHY-1125897, the Kavli Foundation and the Gordon and Betty Moore Foundation grant GBMF 947. 
This research has made use of NASA's Astrophysics Data System.
}

\bibliographystyle{apj}
%\bibliography{../bibtex/spt_smg}

\clearpage
\end{document}